\documentclass[preprint2]{aastex}

\slugcomment{Accepted by Astrophysical Journal}

\shorttitle{Disappearance of Dust Disks}
\shortauthors{Takeuchi, \& Lin}

\begin{document}


\title{Attenuation of Millimeter Emission from Circumstellar Disks
  Induced by the Rapid Dust Accretion
\footnote{Accepted by ApJ}}

\author{Taku Takeuchi\altaffilmark{1,2} and D. N. C. Lin\altaffilmark{2}
}
\altaffiltext{1}{Earth and Planetary Sciences, Kobe University, Kobe 657-8501,
Japan; taku@kobe-u.ac.jp}
\altaffiltext{2}{UCO/Lick Observatory, University of California, Santa Cruz,
CA95064; lin@ucolick.org}

\begin{abstract}
From millimeter observations of classical T Tauri stars, it is 
suggested that dust grains in circumstellar disks have grown to millimeter
size or larger.
However, gas drag on such large grains induces rapid 
accretion of the dust.
We examine the evolution of dust disks composed of 
millimeter sized grains, and show that rapid accretion of the dust disk
causes attenuation of millimeter continuum emission.
If a dust disk is composed mainly of grains of $1 \ {\rm cm}-1$ m, 
its millimeter emission goes off within $10^6$ yr.
Hence, grains in this size range cannot be a main population of the 
dust.
Considering our results together with grain growth suggested by 
the millimeter continuum observations, we expect that the millimeter continuum
emission of disks comes mainly from grains in a narrow size range of
$[1 \ {\rm mm}-1 \ {\rm cm}]$.
This suggests either that growth of millimeter sized grains to centimeter
size takes more than $10^6$ yr, or that millimeter sized grains are 
continuously replenished.
In the former case, planet formation is probably difficult, especially in 
the outer disks.
In the latter case, reservoirs of millimeter grains are possibly large 
($\ga 10$ m) bodies, which can reside in the disk more than $10^6$ yr.
Constraints on the grain growth time-scale are discussed for the above 
two cases.

\end{abstract}



\section{Introduction}


Observations at millimeter wavelengths ($\lambda=1-3$ mm) have revealed
that about 50\% of classical T Tauri stars (CTTSs) emit detectable
millimeter continuum. 
The detection limit, $10-20$ mJy, of the $1.3$ mm observations by
Osterloh \& Beckwith (1995) corresponds to a minimum disk luminosity
of $4 \pi D^2 \nu F_{\nu} \approx 5 \times 10^{28} \ {\rm erg \ s^{-1}}$,
where we used the distance to the Taurus-Auriga molecular cloud, $D=140$
pc, while the typical value of the disk luminosity is $10^{29} -10^{31}
\ {\rm erg \ s^{-1}}$ (Beckwith et al. 1990; Ohashi et al. 1991, 1996).
The sizes of the disks that emit millimeter continuum are in a range
between $100-500$ AU (Dutrey et al. 1996; Kitamura et al. 2002).
Only a small fraction ($\sim 10$\%) of weak line T Tauri stars (WTTSs),
however, have detectable emission at $1.3$ and $2.7$ mm (Osterloh \&
Beckwith 1995; Dutrey et al. 1996; Duvert et al. 2000).
It is not clear whether the typical age of WTTSs is older than that
of CTTSs, as both classes have similar age distributions
between $10^5$ and $10^7$ yr (Kenyon \& Hartmann 1995).
However, the observations have made clear that the outer part ($\ga 100$
AU) of the disk that emits millimeter continuum disappears in a
timescale between $10^6$ and $10^7$ yr.
The standard interpretation assumes that WTTSs are the older counterparts
of CTTSs.

A notable feature of the millimeter emission from CTTSs is the evolution of the
spectral index from the interstellar value. 
The frequency dependence of the dust emissivity (or the opacity)
$\kappa_{\nu}$ is conventionally expressed as a power law,
$\kappa_{\nu} \propto \nu^{\beta}$, where $\nu$ is the frequency of
radio emission. 
Beckwith \& Sargent (1991) argued that the value of the opacity index
$\beta$ evolves from the interstellar value of 2 to $\la 1$ for CTTS
disks (see also Mannings \& Emerson 1994; Hogerheijde et al. 2003).
This change in $\beta$ is probably due to dust evolution in the disks.
However, because different directions of dust evolution (such as in
grain size, in chemical composition, and in grain shape) can
produce observable changes in the dust opacity, it is not easy to
specify the cause of changing $\beta$ (Beckwith, Henning, \& Nakagawa
2000).
Among the above possible causes, dust growth is an
attractive hypothesis in a context of planet formation.
The effect of dust growth on the dust opacity was studied by Miyake \&
Nakagawa (1993).
Their calculations showed that grain growth to 1 mm would produce a
low $\beta$ value ($\la 1$), but that it is difficult to explain $\beta
\la 1$ as long as the maximum size of grains is smaller than $1 \
\micron$.
In this paper, we stand at the ground that the low $\beta$ is a sign of
dust growth, and 
we assume that in the disks showing $\beta \la 1$ the dust grains
have grown to $\ga 1$ mm.

In summary, radio observations have shown that about half of CTTSs
emit detectable millimeter continuum.
Images of millimeter continuum show extended disks of $100-500$ AU
in size, while their wavelength dependences show the
evolution of $\beta$, suggesting grain growth to $\ga 1$ mm.
Millimeter continuum emissions of CTTSs last for $10^6 - 10^7$ yr.

Dust grains in a gas disk migrate to the central star due to gas drag 
(see \S\ref{sec:rad_mig}).
As grains grow larger, their orbital decay time becomes shorter.
When the grain size becomes 1 cm, the decay time is only about $10^5$
yr, which is much shorter than the dust disk's life time observed for
CTTSs.
This dust growth and subsequent dust accretion induce a decline in
millimeter emission from the disk.
The millimeter emission would disappear in $10^5$ yr, if most of the dust
mass had been confined in the largest grains of $\ga 1$ cm.
Thus, we expect that in CTTS disks, there must remain sufficient dust
grains less than 1 cm over $10^6$ yr.
This expectation is confirmed in \S\ref{sec:evo_den} using numerical
calculations.
At the same time, however, the largest grain size must be larger than
about 1 mm, if low $\beta$ is due to dust growth.
Very large bodies (say $\ga 10$ m) experience little migration,
because such bodies almost completely decouple from the gas.
However, such large bodies do not contribute to millimeter emission
(Miyake \& Nakagawa 1993).
Hence, in this paper we conclude that only a narrow range of the dust size,
$[1 \ {\rm mm}, 1 \ {\rm cm}]$, can fulfill the requirement that the
millimeter continuum having $\beta \la 1$ lasts over $10^6$ yr.

Our calculations suggest that the millimeter continuum of CTTS disks is
mainly emitted from grains of $1 \ {\rm mm}- 1 \ {\rm cm}$.
Thus, the lifetime of observable millimeter continuum, $10^6 -10^7$ yr,
constrains the growth time of millimeter sized grains.
One possibility is that millimeter sized grains do not grow to
centimeter size in $10^6$ yr and they hold the millimeter continuum of
$\beta \sim 1$.
Another possibility is rapid growth of grains in a context of planet
formation: grains in CTTS disks have already grown to sizes of
$\ga 10$ m, which is large enough to resist gas drag and to reside in
the disk more than $10^6$ yr.
We consider these possibilities in \S\ref{sec:growth}.


\section{Dust Properties in Disks}

\subsection{Assumption of the Largest Size of Grains}

The observations discussed in the last section suggest that dust
grains in circumstellar disks have grown to at least 1 mm.
It is possible to consider that even at the CTTS stage of $10^6$
yr the dust grains have grown to sizes much larger than 1 mm.
However, the observations have not set any constraint on the largest
size, and thus we know only its lower limit, $\sim 1$ mm.
In order to choose a specific question, we focus on the epoch at
which the largest grains have just grown to $\sim 1$ mm, i.e., we
assume that the largest size is of order of 1 mm.
At such an epoch, the dust mass and the opacity are dominated by the largest
grains ($\sim 1$ mm) as discussed below.
Hence, in the later sections, we consider only a single size
of the largest grains.

\subsection{Dust Radial Migration \label{sec:rad_mig}}

Dust grains migrate inward to the star under the action of gas drag.
The gas experiences a pressure gradient force,
which usually cancels a part of the central star's gravity 
and induces slower rotation than the Keplerian velocity.
Dust grains, whose orbits are hardly affected by gas pressure, are
supposed to rotate with Keplerian velocity, but experiencing headwind.
They lose their angular momentum and spiral inward to the central star (Adachi,
Hayashi, \& Nakazawa 1976; Weidenschilling 1977).

In this paper, we mainly consider dust grains smaller than 1 cm.
For such small grains, gas drag obeys Epstein's law, which is
applicable when the mean free path of the gas molecules is larger than
the grain size (The mean free path is 7 cm at 1 AU and larger at
larger radii in our standard model; see e.g., Nakagawa, Sekiya, \&
Hayashi 1986).
The gas drag force on a spherical grain is given by
\begin{equation}
{\mbox{\boldmath $F$}}_g = - \frac{4}{3} \pi \rho_g s^2  v_t
({\mbox{\boldmath $v$}}_d - {\mbox{\boldmath $v$}}_g ) \ ,
\end{equation}
where $\rho_g$ is the gas density, $s$ is the grain radius, $v_t = (8/
\pi)^{1/2} c$ is the mean thermal velocity of gas molecules, $c$ is
the gas sound speed, and
$\mbox{\boldmath $v$}_d$ and $\mbox{\boldmath $v$}_g$ are 
the velocities of the grain and the gas, respectively.
The stopping time due to gas drag is $t_s = m_d | {\mbox{\boldmath
$v$}}_d - {\mbox{\boldmath $v$}}_g | / |{\mbox{\boldmath $F$}}_g|$,
where $m_d = \case{4}{3}\pi \rho_p s^3$ is the grain mass, and
$\rho_p$ is the physical density of an individual grain.
We define the non-dimensional stopping time normalized by the orbital
time as,
\begin{equation}
T_s = t_s \Omega_{\rm K}= \frac{\rho_p s v_{\rm K}}{\rho_g r v_t} \ ,
\label{eq:stoptime1}
\end{equation}
where $r$ is the distance from the star, $\Omega_{\rm K} =
(GM/r^3)^{1/2}$ is the Keplerian angular velocity, $v_{\rm K} = r
\Omega_{\rm K}$ is the Keplerian velocity, and $M$ is the central star's
mass.

\begin{figure}
\epsscale{1.0}
\plotone{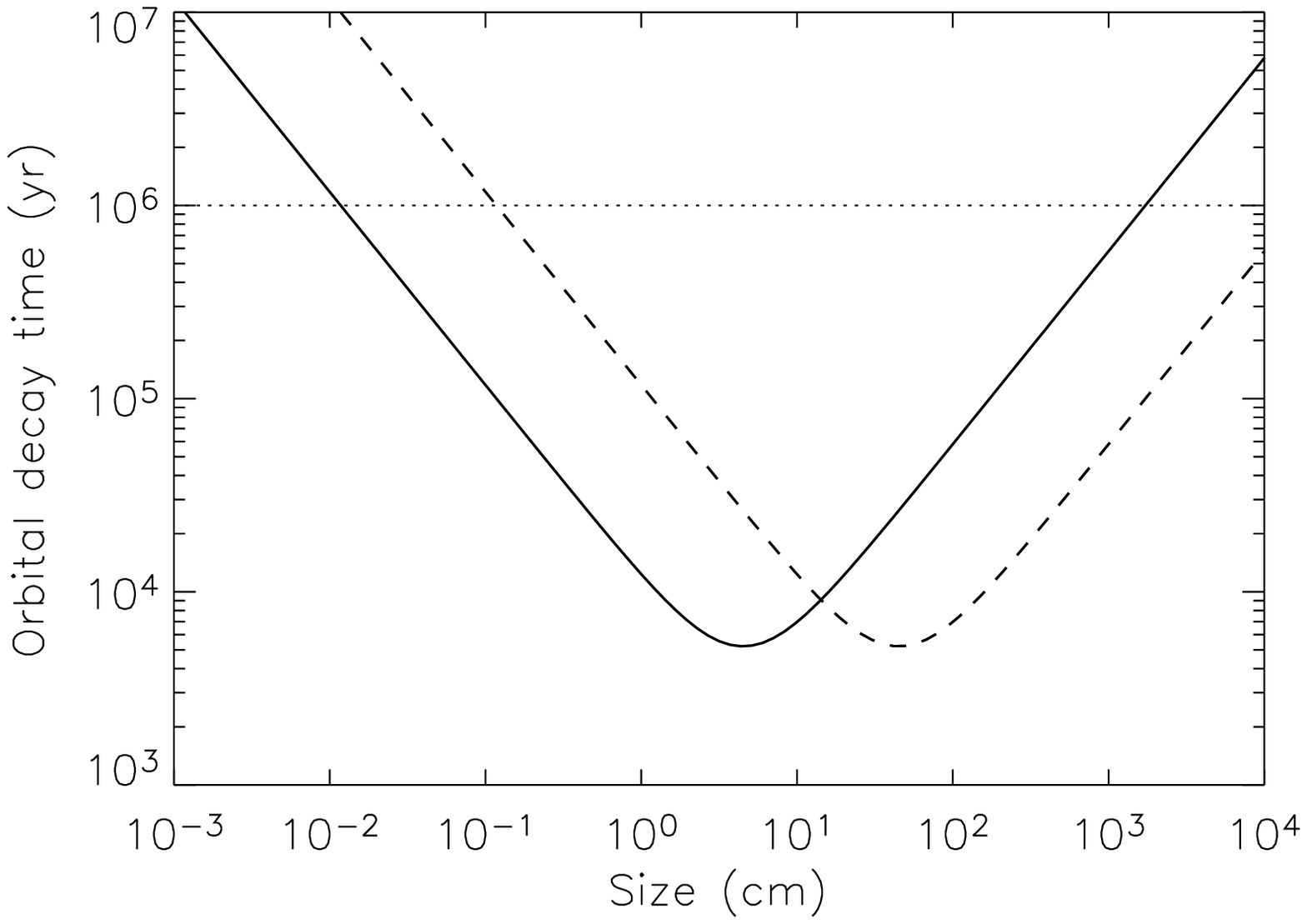}
\caption{
Orbital decay time, $\tau_{\rm dust}$, of dust grains at 50 AU in a gas disk.
The solid line represents the orbital decay time of compact grains of
physical density $\rho_p = 1 \ {\rm g \  cm}^{-3}$, and the dashed line
shows that of fluffy grains of $\rho_p = 0.1 \ {\rm g \ cm}^{-3}$.
The decay time is calculated at the midplane of the model disk
described in \S\ref{sec:diskmodel}, but ignoring the gas accretion
velocity ($v_g=0$).
The dotted line represents a reference time scale of $10^6$ yr.
\label{fig:driftvel}
}

\end{figure}

The stopping time is proportional to the grain size, which means that
the gas drag force is more effective for smaller grains.
If a grain is small enough to have a stopping time much less than its
orbital time ($T_s \ll 1$; this is true for grains of $s \la 1$ cm), it
orbits with the almost same angular velocity as the gas has, which is
sub-Keplerian (Adachi et al. 1976; Weidenschilling 1977).
Therefore, the centrifugal force on the grain is smaller than the
stellar gravity, resulting in an inward acceleration by the residual gravity.
This acceleration is however balanced by gas drag in the radial
direction.
After a stopping time-scale, which is smaller than the orbital time, the
grain reaches a terminal falling velocity.
This terminal velocity is slower for smaller grains because the stronger
gas drag force per unit mass holds the grains more tightly with the gas.
For large grains of $T_s > 1$, the coupling between the grains
and the gas is weak.
They rotate with Keplerian velocity and experience headwind from the gas.
The inward drift speed of the grains is proportional to the angular
momentum loss by gas drag ($\propto s^2$ ) per unit mass ($\propto s^2 /
s^3$).
Thus, larger grains drift more slowly for $T_s > 1$ regime.
Figure \ref{fig:driftvel} shows the orbital decay time (or the migration
time $\tau_{\rm dust}$ defined in \S\ref{sec:timescale}) of grains at 50
AU in a disk model described below in \S\ref{sec:diskmodel}.
(Epstein's law is valid for $s \la 1$ km at 50 AU. We used
eq. [\ref{eq:dustvel1}] described below to calculate $\tau_{\rm
dust}=r/|v_d|$, assuming the gas accretion velocity $v_g=0$.)
For compact grains (the particle physical density $\rho_p = 1 \ {\rm g \
cm}^{-3}$; {\it solid line}), the orbital decay time of $[100 \ \micron
- 10 \ {\rm m}]$ grains is less than $10^6$ yr, and it has a minimum
value of $5 \times 10^3$ yr when the size is 5 cm.
For fluffy grains ($\rho_p = 0.1 \ {\rm g \ cm}^{-3}$; {\it dashed
line}), the plot of orbital decay time in Figure \ref{fig:driftvel}
shifts rightward with a 10 fold increase in the grain size.
From this figure, it is expected that grains of $\rho_p = 0.1-1 \ {\rm g
\ cm}^{-3}$ in the size range $[1 \ {\rm mm} - 10 \ {\rm m}]$ disappear
from the disk within $10^6$ yr because of rapid accretion to the central
star.

\subsection{Dust Opacity \label{sec:opacity}}

We focus on the optical property of the dust at millimeter wavelengths,
$\lambda \sim 1$ mm.
As we assumed that the maximum size is of order of 1 mm, most grains
are similar to or smaller than the observation wavelength, $\lambda \sim 1$
mm.
The optical properties of such grains can be described according to
Rayleigh scattering theory. 
The absorption cross section, $\sigma_{\rm abs}$, of a spherical
grain is proportional to its volume or mass (van de Hulst 1981), i.e.,
\begin{equation}
\sigma_{\rm abs} = - \frac{4 \pi m^{\prime}}{\lambda \rho_p} m_d \ ,
\end{equation}
where $m^{\prime}$ is the imaginary part of the refractive index,
$\lambda$ is the wavelength, and $m_d=\case{4}{3} \pi s^3 \rho_p$ is the
grain mass.
The cross section per unit mass is independent of the grain size:
$\sigma_{\rm abs} / m_d = - 4 \pi m^{\prime} / (\lambda \rho_p)$.

Suppose a power-law grain size distribution.
The number density of grains in a unit size range is
\begin{equation}
n(s) = n_0 \left( \frac{2 \pi s}{\lambda} \right)^{-\delta} \ .
\end{equation}
The total cross section of grains in a unit volume is
\begin{eqnarray}
\sigma_{\rm tot} & = &\int_{s_{\rm min}}^{s_{\rm max}} \sigma_{\rm abs}
n (s) ds \nonumber \\
& = & -\frac{8 \pi}{3 (4-\delta)} \left( \frac{2 \pi}{\lambda}
\right)^{1-\delta} n_0 m^{\prime} (s_{\rm max}^{4-\delta} - s_{\rm
min}^{4-\delta}) \ \ \ \ \ \,
\end{eqnarray}
where $s_{\rm min}$ and $s_{\rm max}$ are the minimum and maximum
sizes of grains.
If $\delta$ is less than 4, the cross section is dominated by the
largest grains.
In such a case, it is adequate to take only the largest grains in
consideration of the dust opacity.

When the largest grains have grown to sizes larger than $\sim 10$ m,
it is expected that the dust size distribution of the disk has a void
between $[1 \ {\rm mm} - 10 \ {\rm m}]$, because of rapid accretion of
grains to the star in this size range, as discussed in \S\ref{sec:rad_mig}.
In addition, grains larger than 10 m hardly contribute to
millimeter emission.
The mass opacity of 10 m grains is $10^{-3}$ times smaller
than 1 mm grains (Fig. 4 in Miyake \& Nakagawa 1993).
If most of the dust mass is concentrated in bodies larger than 10 m,
the dust cannot emit a millimeter flux strong enough to explain the
observed values of CTTSs.
Hence, in consideration of the millimeter emission, the contribution of
bodies larger than 10 m can be neglected.

\subsection{Dust Growth \label{sec:th_growth}}

The size distribution of grains in the inter-stellar medium is
considered to be a power-law with $\delta=3.5$ (Mathis, Rumpl, \&
Nordsieck 1977).
In circumstellar disks, however, the grains stick together and grow
larger.
The size distribution probably changes through this process.
Mizuno, Markiewicz, \& V\"olk (1988) found from their numerical
simulations that after the grain growth more mass concentrates in the
largest grains and that the power-law index becomes as small as
$\delta=1.9$.
On the other hand, collisional destruction of grains may lead to
a different distribution.
Hellyer (1970) analytically derived a power-law index $\delta \approx
3.5$ for particles experiencing collisional fragmentation.
While this is in agreement with the observed distribution of asteroids
of sizes between $30-300$ km, recent observations show a shallower
index for small asteroids of $\la 5$ km (Yoshida et al. 2003).
In either case, the largest bodies dominate the dust mass and the
dust opacity at $\lambda \sim 1$ mm, as long as the largest size is less
than $\sim 1$ mm.

In conclusion, the above considerations in \S\ref{sec:opacity} and
\S\ref{sec:th_growth} suggest focusing on the largest grains of
$\sim 1$ mm.
In the following sections, we investigate evolution of the dust density
profile and the thermal flux emitted at millimeter wavelengths, both of
which are dominated by the largest grains.


\section{Model Equations and Assumptions}

As discussed in the last section, we consider disks at the epoch in
which the largest grains are of order of 1 mm, and focus on the
largest grains. 
We solve for the density evolution of a dust disk, assuming all grains are
of a similar size of $\sim 1$ mm.
Contributions from smaller grains are neglected.
In the numerical calculations, the grain size is assumed to be
constant with time and any effect caused by dust growth is ignored.
From the numerical results, however, we discuss the constraint on the
dust growth time-scale in \S\ref{sec:growth} below. 

\subsection{Model Assumptions and the Initial Disk \label{sec:diskmodel}}

For simplicity, we assume that the gas disk has a power-law temperature
profile $T$ in the radial direction $r$, and is isothermal in the vertical
direction $z$.
The profile is written as
\begin{equation}
T(r)=T_0 r_{\rm AU}^q \ ,
\end{equation}
where the subscript ``0'' denotes quantities at 1 AU, and a
non-dimensional quantity $r_{\rm AU}$ is the radius in AU.
We assume that the temperature profile does not change through the disk
evolution.
The initial surface density distribution $\Sigma_g$ is also a power-law,
\begin{equation}
\Sigma_g (r,t=0) = \Sigma_{g,0} r_{\rm AU}^p \ ,
\label{eq:gassurface}
\end{equation}
and is truncated at $r_{\rm out}=100$ AU.
The isothermal sound speed is $c=c_0 r_{\rm AU}^{q/2}$ and the gas
disk scale height $h_g$ is thus defined as
\footnote{Strictly speaking,
the scale height of an isothermal disk is $\sqrt{2} h_g$, because the
gas density varies as $\rho_g \propto \exp [-z^2/(2 h_g^2)]$.}
\begin{equation}
h_g(r) \equiv \frac{c}{\Omega_{\rm K}} = h_0 r_{\rm AU}^{(q+3)/2} \ .
\label{eq:gasheight}
\end{equation}
The disk has a turbulent viscosity $\nu_t$.
In this paper, we simply model the viscous effect of turbulence using the
so-called $\alpha$ prescription (Shakura \& Sunyaev 1973),
\begin{equation}
\nu_t = \alpha c h_g = \alpha c_0 h_0 r_{\rm AU}^{q+(3/2)} \ .
\label{eq:alphavis}
\end{equation}

Most of the dust mass is concentrated in the largest grains.
Therefore, we assume all grains have the same size, which is of order
of 1 mm.
Initially, the dust spreads over the entire gas disk, and its density
profile $\Sigma_d$ is proportional to the gas density:
\begin{equation}
\Sigma_d (r,t=0) = f_{\rm dust} \Sigma_g = \Sigma_{d,0} r_{\rm AU}^p \ ,
\label{eq:dustsurface}
\end{equation}
where the initial dust-to-gas ratio $f_{\rm dust}$ is independent of $r$.
Most of the disk is outside the ice condensation radius ($\sim 3$
AU), and the main composition of the dust is water ice.
We ignore sublimation of water ice at the innermost part of the disk,
because our interest is on the outer part that contributes to
millimeter emission.
The physical density of a compact grain is assumed to be $\rho_p = 1 \ {\rm
g \ cm}^{-3}$.
If grains are fluffy (or porous) and include vacuum inside their
volume, the density is smaller.

We adopt the following fiducial parameters:
$M = 1 \ M_\sun$, $\Sigma_{g,0} = 3.5 \times 10^2 \ {\rm g \ cm}^{-2}$,
$h_0 = 3.33 \times 10^{-2}$ AU, $p=-1$, $q=-\case{1}{2}$, $r_{\rm out} = 100$
AU, and $\alpha=10^{-3}$.
These values correspond to a disk having a gas mass of 
$M_g = 2.5 \times 10^{-2} M_{\sun}$ inside 100 AU (the mass within 40 AU
is $10^{-2} M_{\sun}$, which is comparable to the
minimum-mass-solar-nebula model of the standard theory of planet
formation; Hayashi, Nakazawa, \& Nakagawa 1985).
The temperature distribution is $T= 278 \ r_{\rm AU}^{-1/2} \ {\rm K}$
(Hayashi et al. 1985).
Radio observation by Kitamura et al. (2002) suggests that
the power law index of the surface density profile, $p$, is between $-1.5
< p < 0$.
For dust grains, we use $f_{\rm dust} = 10^{-2}$, $s = 1$ mm, and
$\rho_p = 1 \ {\rm g \ cm}^{-3}$.

\subsection{Equations for the Density Evolution}

Evolution of a gas disk occurs due to viscous torque
(Lynden-Bell \& Pringle 1974).
The torque exerted on the disk gas outside $r$ by the inner disk is $-2
\pi \nu_t \Sigma_g r^3 d\Omega_{\rm K} / dr$.
The differential between the torques exerted from the inside and from the
outside causes a radial motion of the gas.
Assuming that the gas rotation velocity is always Keplerian, the gas
radial velocity becomes
\begin{equation}
v_g = - \frac{3}{r^{1/2} \Sigma_g} \frac{d}{dr} ( r^{1/2} \nu_t \Sigma_g)
\ ,
\label{eq:gasvel}
\end{equation}
as derived in equation (12) of Lin \& Papaloizou (1986; with no
planetary torque).
The gas density $\Sigma_g$ obeys the equation of continuity,
\begin{equation}
\frac{\partial}{\partial t} \Sigma_g - \frac{3}{r}
\frac{\partial}{\partial r} \left[ r^{1/2} \frac{\partial}{\partial r} (
r^{1/2} \nu_t \Sigma_g) \right] = 0 \ .
\label{eq:cont_gas}
\end{equation}

The equation of continuity for the dust density $\Sigma_d$ is
\begin{equation}
\frac{\partial}{\partial t} \Sigma_d + \frac{1}{r}
\frac{\partial}{\partial r} \left[ r ( F_{\rm dif} + \Sigma_d v_d )
  \right] = - \dot{\Sigma}_d \ ,
\label{eq:cont_dust}
\end{equation}
where $F_{\rm dif}$ is the diffusive mass flux, $v_d$ is the dust
radial velocity, and $ \dot{\Sigma}_d$ is the mass loss rate of the dust
due to grain growth.
The mass flux comes from two terms.
One is the mass flux caused by diffusion of grains in the
turbulent gas.
The diffusion works to smooth out any fluctuation in the dust concentration
$\Sigma_d / \Sigma_g$.
We adopt the assumption that the diffusive mass flux is proportional to
the concentration gradient, and that the diffusion coefficient is the same
as the gas viscosity.
This assumption is applicable to small passive grains tightly coupled
to the gas ($T_s \ll 1$ and the Schmidt number ${\rm Sc}=1$).
The diffusive mass flux $F_{\rm dif}$ is defined by
\begin{equation}
F_{\rm dif} = - \nu_t \Sigma_g \frac{\partial}{\partial r} \left(
  \frac{\Sigma_d}{\Sigma_g} \right) \ .
\end{equation}

The other contribution to the mass flux is from the dust radial motion
induced by gas drag and is written as $\Sigma_d v_d$.
The dust radial velocity, $v_d$, caused by gas drag is calculated in
equation (23) of Takeuchi \& Lin (2002; henceforth Paper I) as
\begin{equation}
v_d = \frac{T_s^{-1} v_g - \eta v_{\rm K}}{T_s + T_s^{-1}} \ ,
\label{eq:dustvel1}
\end{equation}
where $\eta$ is a factor relating the gas rotation velocity to
its Keplerian value.
The gas disk has a radial pressure gradient, $\partial P_g / \partial
r$, which cancels a part of the stellar gravity and leads to a slower
gas rotation than the Keplerian velocity.
We define $\eta$ as the ratio of the pressure gradient to the gravity,
\begin{equation}
\eta = -\frac{1}{r \Omega_{\rm K}^2 \rho_g} \frac{\partial P_g}{\partial
  r} \ .
\label{eq:eta1}
\end{equation}
Because the gravity is reduced by a factor ($1-\eta$), the angular
velocity of the gas becomes
\begin{equation}
\Omega_g^2 (r) = \left( 1 - \eta \right) \Omega_{\rm K}^2(r) \ .
\label{eq:veldiff}
\end{equation}
As discussed in Paper I, the gas rotation velocity and $\eta$
are functions of the height $z$ from the midplane.
The stopping time $T_s$ also varies because the gas density
changes with $z$.
In the calculation of the dust velocity in equation (\ref{eq:dustvel1}),
we use the midplane values for $\eta$ and $T_s$.
This is because grains as large as 1 mm sediment to the midplane and
most grains are concentrated at the midplane.
(Concentration of 1 mm grains is not strong enough to induce a change
in the gas velocity profile at the midplane, provided that $\alpha \ga
10^{-4}$. The midplane dust density does not exceed the gas density, as
discussed in \S\ref{sec:fastgrowth} below.
We use the gas radial velocity of eq. [\ref{eq:gasvel}], which is
the averaged value over the $z$-direction.
The outward motion of the midplane gas proposed by Paper I is neglected,
because it has a significant effect only at the innermost part ($r \la
3$ AU) in which the first term of eq. [\ref{eq:dustvel1}] dominates the dust
velocity.)
At the midplane, the gas density is (eq. [6] of Paper I)
\begin{equation}
\rho_g = \frac{\Sigma_g}{\sqrt{2 \pi} h_g} \ .
\label{eq:midgasden}
\end{equation}
The pressure is calculated from the isothermal equation
of state as 
\begin{equation}
P_g = c^2 \rho_g = \frac{h_g \Omega_K^2
  \Sigma_g}{\sqrt{2 \pi}} \ .
\end{equation}
Using the power-law expression of the disk scale height
(\ref{eq:gasheight}) and the Kepler rotation law $\Omega_{\rm K} \propto
r^{-3/2}$, we have the midplane value
\begin{equation}
\eta = -\left( \frac{h_g}{r} \right)^2 \left( \frac{d \ln 
  \Sigma_g}{d \ln r} + \frac{q-3}{2} \right) \ .
\label{eq:eta2}
\end{equation}
Similarly, using equations (\ref{eq:stoptime1}), (\ref{eq:gasheight})
and $v_t=\sqrt{8/\pi} c$, the non-dimensional stopping time at the
midplane is
\begin{equation}
T_s = \frac{\pi \rho_p s}{2 \Sigma_g} \ .
\label{eq:stoptime2}
\end{equation}

In the right-hand-side in equation (\ref{eq:cont_dust}), a sink term
$\dot{\Sigma}_d$ is added to represent a mass loss of the dust.
The mass loss is caused by grain growth and subsequent rapid
accretion to the star.
Detailed definition of the mass loss is discussed in \S\ref{sec:growth}
below.

\subsection{A Convenient Non-dimensional Parameter}

In a gas disk with given profiles of $\Sigma_g$ and $\eta$, the dust
density evolution is controlled only via the stopping time $T_s$ (see
eqs. [\ref{eq:cont_dust}] and [\ref{eq:dustvel1}]).
If different sets of the dust parameters $\{s, \rho_p \}$ have the same
stopping time, evolutions of these dust disks are the same (we assume
the sink term $\dot{\Sigma}_d=0$).
In addition, scaling the gas density does not change the gas evolution (see
eq. [\ref{eq:cont_gas}]).
That is, if the initial gas mass (or $\Sigma_{g,0}$ in
eq. [\ref{eq:gassurface}]) is altered without changing the shape of
the density profile ($p$ in eq. [\ref{eq:gassurface}]), we have the same
gas evolution except for the scaling factor.
Throughout this paper, we consider only the case of $\Sigma_g \propto
r^{-1}$ and $r_{\rm out}=100$ AU.
The dust evolution depends also on the initial gas mass through the
stopping time (eq. [\ref{eq:stoptime2}]).
However, we can retain the same dust evolution for different values of
the initial gas mass by adjusting the dust parameters to keep the same
stopping time.
For parameter sets of the gas mass $M_g$ within 100 AU, the grain
size $s$, and the particle physical density $\rho_p$, if we have the
same value of
\begin{equation}
A \equiv \left( \frac{M_g}{2.5 \times 10^{-2} M_{\sun}} \right)^{-1} 
\left( \frac{s}{1 \ {\rm mm}} \right)
\left( \frac{\rho_p}{1 \ {\rm g \ cm}^{-3}} \right) \ ,
\label{eq:A1}
\end{equation}
then we have the same stopping time and the same dust evolution.
(If the initial profile of the gas surface density is different from
$\Sigma_g \propto r^{-1}$, we have different dust evolution even for the
same value of $A$.)

For example, the parameter sets $\{M_g,s, \rho_p \} = \{2.5 \times 10^{-2}
  \ M_{\sun}, 1 \ {\rm mm}, 1 \ {\rm g \ cm}^{-3} \}$, \
$\{2.5 \times 10^{-2} \ M_{\sun}, 10 \ {\rm mm}, 0.1   \ {\rm g \
  cm}^{-3} \}$, \
and $\{2.5 \times 10^{-1} \ M_{\sun}, 10 \ {\rm mm}, 1 \ {\rm g \
  cm}^{-3} \}$
have the same $A=1$.
We call these parameter sets ``$A=1$ model''.
The other parameter sets considered in this paper have $A=0.1$ or 10. 
The parameter sets of $A=0.1$ model are, for example,
$\{M_g,s, \rho_p \} = \{2.5 \times 10^{-1}
  \ M_{\sun}, 1 \ {\rm mm}, 1 \ {\rm g \ cm}^{-3} \}$ or
$\{2.5 \times 10^{-2} \ M_{\sun}, 1 \ {\rm mm}, 0.1   \ {\rm g \
  cm}^{-3} \}$.
In $A=10$ model, we have for example
$\{M_g,s, \rho_p \} = \{2.5 \times 10^{-3}
  \ M_{\sun}, 1 \ {\rm mm}, 1 \ {\rm g \ cm}^{-3} \}$ or
$\{2.5 \times 10^{-2} \ M_{\sun}, 10 \ {\rm mm}, 1   \ {\rm g \
  cm}^{-3} \}$.

We note that the millimeter continuum flux from the dust disk depends on
the disk mass (nearly proportional to the total dust mass).
When calculating millimeter flux, we specify the gas disk mass $M_g$, in
addition to the value of $A$.
If the dust sink term $\dot{\Sigma}_d$ is considered, the disk evolution
varies independently with $s$, $\rho_p$, and $M_g$.

Figure \ref{fig:stoptime} shows the non-dimensional stopping time for
various values of $A$.
The non-dimensional stopping time is less than unity in most cases of
our models.
Grains are tightly coupled to the gas except at the outermost part
of the disk of $A=10$.

\subsection{Accretion Time Scale of the Gas and the Dust \label{sec:timescale}}

The gas accretion time-scale $\tau_{\rm gas}$ is of order of the viscous
time-scale $r^2/ \nu_t$.
Here, we define it as $\tau_{\rm gas} = r / |v_g|$, where the
gas radial velocity is given by equation (\ref{eq:gasvel}).
From equation (\ref{eq:alphavis}), the viscosity $\nu_t$ is proportional to
$r$ in the models with $q=-\case{1}{2}$, and then the accretion time $\tau_{\rm
gas}$ is also proportional to $r$, as shown by the dashed line in
Figure \ref{fig:time}.
The accretion time is of order of $10^7$ yr at 100 AU when the viscosity
parameter is $\alpha = 10^{-3}$.
This means that disks with an initial radius of 100 AU keep their gas
for $10^7$ yr.

\begin{figure}
\epsscale{1.0}
\plotone{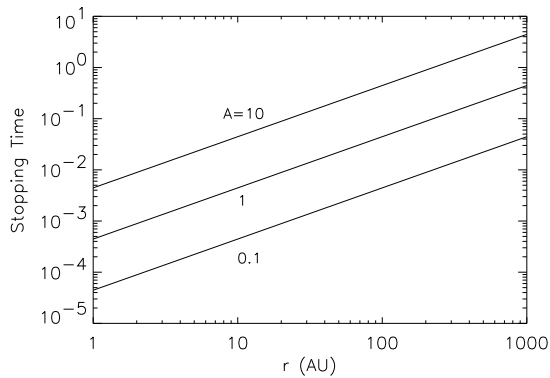}
\caption{
Non-dimensional stopping time of a dust grain for various models.
From the upper line, $A=10$, 1, and $0.1$, respectively.
Model $A=1$ is the fiducial model, which have for example $M_g =2.5
\times 10^{-2} \ M_{\sun}$, $\rho_p = 1 \ {\rm g \ cm}^{-3}$, and $s=1$ mm.
Model $A=0.1$ represents for example a lower physical density of a grain
($\rho_p = 0.1 \ {\rm g \  cm}^{-3}$) or a higher gas disk mass ($M_g
=0.25 M_{\sun}$).
Model $A=10$ represents for example a larger grain size ($s=1$ cm).
In this figure, the power-law gas density profile is extrapolated to
1000 AU.
The gas mass $M_g$ represents the mass within 100 AU.
\label{fig:stoptime}
}
\end{figure}

\begin{figure}
\epsscale{1.0}
\plotone{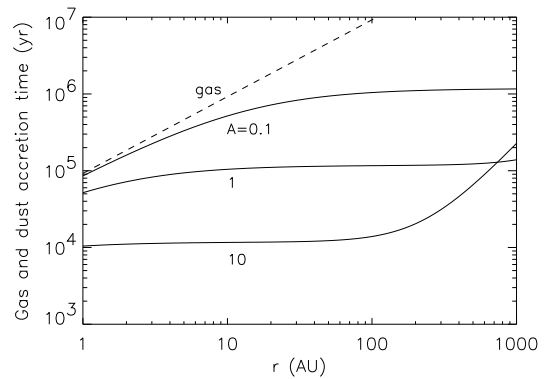}
\caption{
Accretion time-scales of the gas and the dust in the fiducial model
defined in \S\ref{sec:diskmodel}.
The dashed line shows the gas accretion time-scale, $\tau_{\rm gas}$,
calculated for $\alpha=10^{-3}$.
The solid lines show the dust migration time-scales, $\tau_{\rm dust}$,
of the models for $A=0.1$, 1, and 10, respectively, from the upper line.
In all models, the gas accretion time $\tau_{\rm gas}$ is the same.
The power-law gas density profile is extrapolated to 1000 AU.
\label{fig:time}
}
\end{figure}

The inward migration time of the dust is defined by $\tau_{\rm dust} = r
/|v_d|$, where the dust radial velocity written in equation
(\ref{eq:dustvel1}) is induced by gas drag.
The stopping time due to gas drag is smaller than the orbital time in
most cases considered here, i.e., $T_s <1$, and grains are
tightly coupled to the gas, as shown in Figure \ref{fig:stoptime}.
From equations (\ref{eq:dustvel1}) and (\ref{eq:stoptime2}), the radial
velocity of small grains that have $T_s \ll 1$ becomes 
\begin{equation}
v_d = v_g - \frac{\pi \eta \rho_p s v_{\rm K}}{2 \Sigma_g} \ ,
\label{eq:dustvel2}
\end{equation}
which means that the radial drift velocity from the gas (the second term) is
proportional to the grain size and the particle physical density, and is
inversely proportional to the gas density.

Figure \ref{fig:time} shows the dust migration time $\tau_{\rm dust}$
for various models.
Model of $A=1$ represents, for example, 1 mm grains with $\rho_p = 1 \ {\rm
  g \ cm}^{-3}$ in a disk of a mass $M_g=2.5\times 10^{-2} M_{\sun}$
inside 100 AU.
In this case, the migration time is about $10^5$ yr and does not vary
significantly with $r$.
The time-scale is much shorter than the gas accretion time-scale in most part
of the disk.
Hence, we expect such grains are quickly accreted onto the star and
removed from the disk.
The migration time is even smaller for larger grains or for a smaller
gas mass in $A=10$ model.
Outside 200 AU, the migration time increases with $r$, because the
non-dimensional stopping time is larger than unity at such a region (see
Fig. \ref{fig:stoptime}), and the dust velocity decreases with
decreasing $\Sigma_g$
(see eqs. [\ref{eq:dustvel1}] and [\ref{eq:stoptime2}]).
In $A=0.1$ model, i.e., if grains are fluffy ($\rho_p = 0.1 \ {\rm
  g \ cm}^{-3}$) or if the gas disk is massive ($M_g=2.5 \times 10^{-1}
M_{\sun}$), then the dust migration time can be as long as $10^6$ yr in
the outer part of the disk ($r \ga 50$ AU).


\section{Evolution of the Dust Distribution \label{sec:evo_den}}

We first show how the radial motion of the dust accelerates
evolution of the dust density and of its millimeter emission.
In this section, the sink term due to grain growth in equation
(\ref{eq:cont_dust}) is set to zero, i.e., $\dot{\Sigma}_d = 0$.
The effect of grain growth is considered in \S\ref{sec:growth} below.

\subsection{Numerical Method}

We solve equations (\ref{eq:cont_gas}) and (\ref{eq:cont_dust})
numerically in order to obtain evolution of the density profiles of the gas
and the dust.

Equation (\ref{eq:cont_gas}) for the gas is a diffusion equation.
We solve the equation using the standard implicit method, the
Cranck-Nicholson scheme (Press et al. 1992).
The numerical mesh is uniformly spaced in a logarithmic space of
$[r_1, r_2]=[1,  10^3]$ AU by 200 grid points.
For boundary conditions, we simply assume that the density at the inner
and outer boundaries are zero, i.e.,
$\Sigma_g(r_1)=\Sigma_g(r_2)=0$ (Bath \& Pringle, 1981; Armitage et
al. 2003).

For the dust, equation (\ref{eq:cont_dust}) is composed of the terms of
diffusion, $\partial (r F_{\rm dif}) / \partial r$, and of advection,
$\partial (r \Sigma_d v_d) / \partial r$.
Each contribution of these terms to the time evolution is numerically
solved in separate steps.
The diffusion part is solved with the Cranck-Nicholson scheme.
The advection part is solved with the monotonic second order scheme (van
Leer 1977).
We use the same boundary conditions as the gas, i.e.,
$\Sigma_d(r_1)=\Sigma_d(r_2)=0$. 

The inner boundary condition, $\Sigma_g(r_1)=0$, induces an artificial
positive density gradient on the close neighborhood of the boundary.
We regard that the evolution of the main body of the gas disk will not be
affected seriously by this boundary condition.  
For the dust, however, the positive gas density gradient leads to a
positive gas pressure gradient and thus $\eta < 0$, which raises an
artificial outward motion of the dust (see eqs. [\ref{eq:dustvel1}] and
[\ref{eq:eta1}]).
To suppress this outward motion, we simply set $\eta =
-(h_g/r)^2(q-3)/2\,(>0)$, neglecting the term 
$d (\ln \Sigma_g) / d (\ln r)$ in equation (\ref{eq:eta2}), in the
inner-edge zone where the gas density gradient is positive.

\subsection{Low Viscosity Models \label{sec:lowvis}}

First, we discuss disks with a low viscosity.
When the viscosity parameter is $\alpha=10^{-3}$, the gas accretion
time-scale at the outer disk radius (100 AU) is $10^7$ yr.
The gas disk survives viscous draining for $10^7$ yr.
Grains of $s=1$ mm and $\rho_p = 1 \ {\rm g \ cm}^{-3}$, however, have
migration time as small as $10^5$ yr, and accrete much more quickly to
the star.

\begin{figure}
\epsscale{1.0}
\plotone{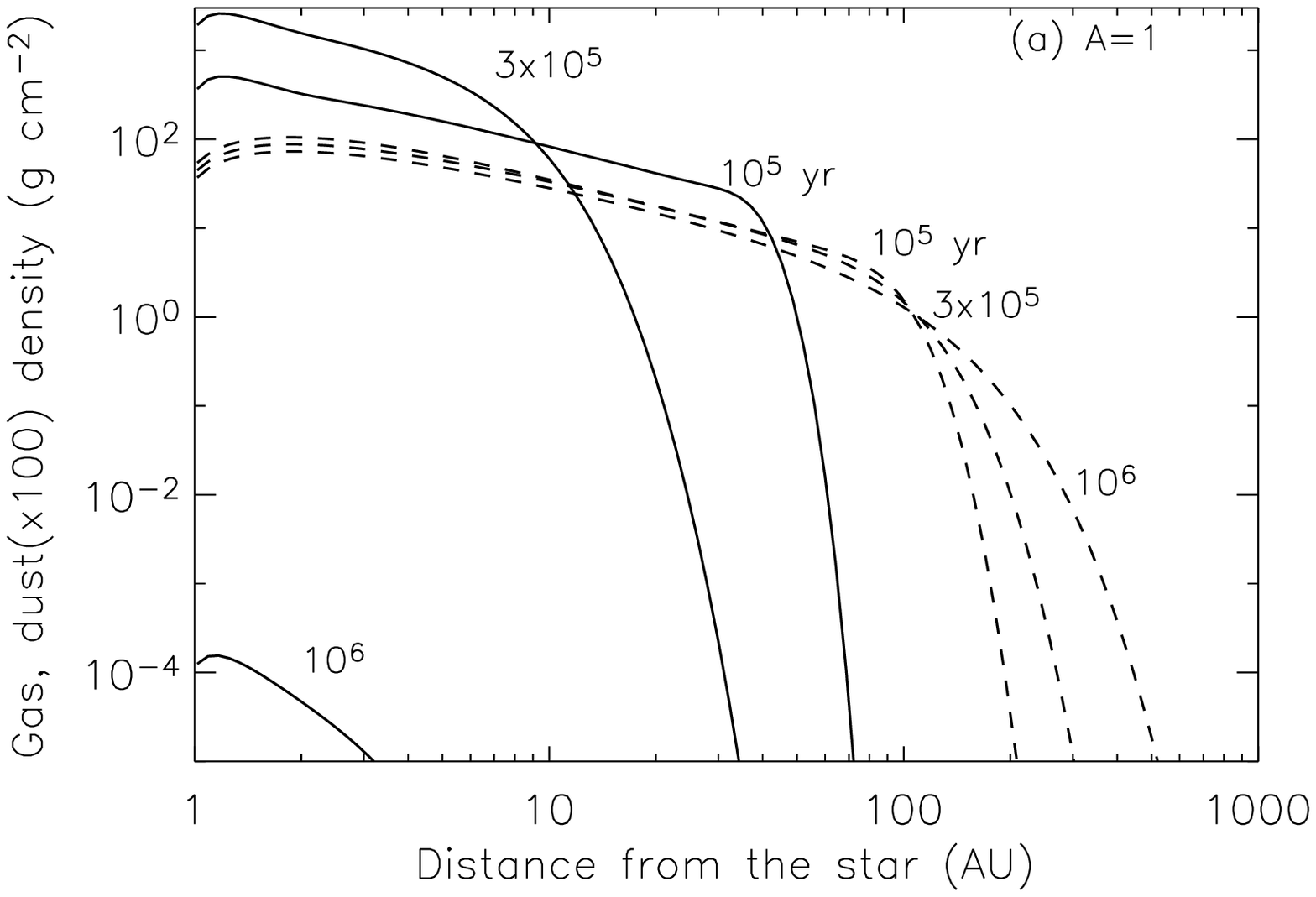}
\plotone{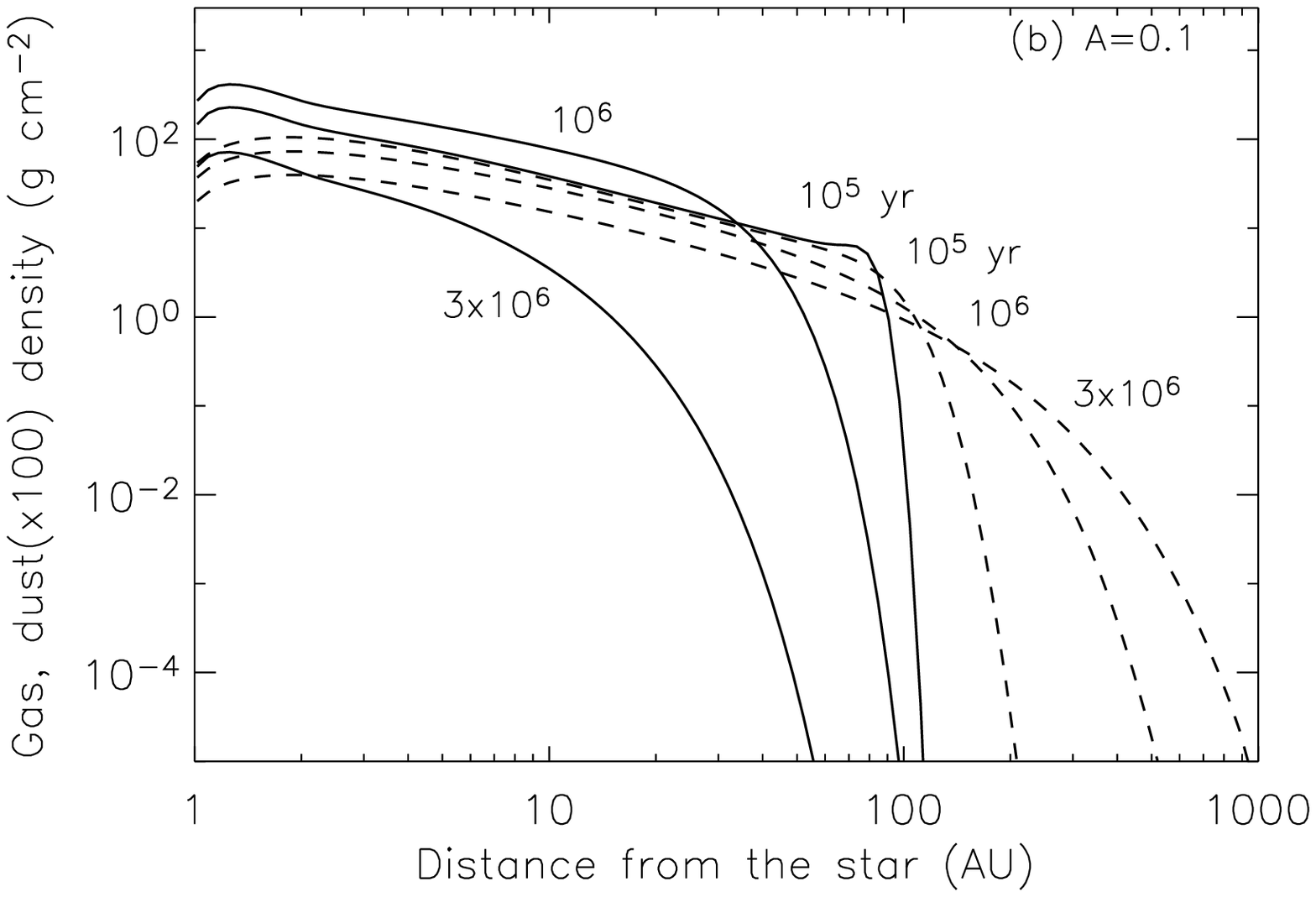}
\caption{
Evolution of the gas and dust surface densities of a low viscous disk
($\alpha=10^{-3}$).
The values of the dust surface densities are multiplied by $f_{\rm
dust}^{-1}=100$ and shown by the solid lines.
The dashed lines show the gas surface densities.
 (a) $A=1$, i.e., for example,
 $M_g =2.5 \times 10^{-2} \ M_{\sun}$, $\rho_p = 1 \ {\rm g \ cm}^{-3}$,
and $s=1$ mm. 
(b) $A=0.1$, i.e., for example, $\rho=0.1 \ {\rm g \  cm}^{-3}$ or
$M_g=0.25 M_{\sun}$.
(In the $M_g=0.25 M_{\sun}$ case, the labels of the $y$-axis must be
read as 10 times larger values.)
\label{fig:denevo}
}
\end{figure}

Figure \ref{fig:denevo}$a$ shows the surface density evolution of the gas
and the dust in $A=1$ model.
While the outermost part of the gas disk expands, the inner part accretes 
inward.
The gas surface density in a region $[1, 100]$ AU slowly decreases
from the initial value.
Although the dust disk has the same initial density profile
as the gas except for the dust-to-gas ratio $f_{\rm dust}$,
it experiences quite a different evolution.
The inward drift velocity is large, and none of the
dust can expand beyond the initial outer radius of 100 AU.
The entire dust disk rapidly accretes to the star.
In $10^6$ yr, most of the dust has fallen to the star.
A dust disk of $A=1$ can not survive for $10^6$ yr.

The expression of the dust radial velocity (\ref{eq:dustvel2}) shows
that the dust has a slower velocity if the particle physical density is
smaller or if the gas density is higher.
In such cases, the gas drag force is more effective at keeping the
grains tightly entrained to the slow, gas accreting motion.
The dust density evolution of $A=0.1$ model is 10 times slower than
$A=1$ model, as shown in Figure \ref{fig:denevo}$b$.
The dust disk can survive over $10^6$ yr in this case.

When the grains grow larger, however, the dust returns to a faster
migration velocity, as shown in equation (\ref{eq:dustvel2}).
Even in the cases where the dust is fluffy ($\rho= 0.1 \ {\rm g \
  cm}^{-3}$) or the gas disk is massive ($M_g = 0.25 M_{\sun}$), if the
grain radius is 1 cm, then we have again $A=1$ and the
dust accretes rapidly in $10^5$ yr.
The dust density evolution goes back to $A=1$ model shown in
Figure \ref{fig:denevo}$a$.

\begin{figure}
\epsscale{1.0}
\plotone{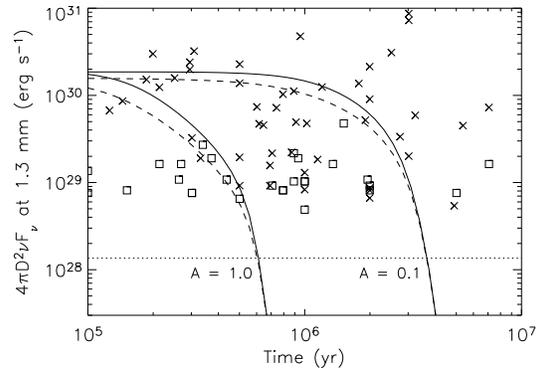}
\caption{
Evolution of the dust thermal emission at a wavelength of $1.3$ mm for a low
viscous disk ($\alpha=10^{-3}$) viewed face-on ($i=0^{\circ}$).
The solid lines are for models of $A=1$ and $0.1$, respectively, from
the left.
The initial disk mass is $M_g=2.5 \times 10^{-2} M_{\sun}$.
The dashed lines show the fluxes of inclined ($i=60^{\circ}$) disks for
comparison. 
Crosses are the observed values of T Tauri stars taken from Beckwith et
al. (1990) and Osterloh \& Beckwith (1995).
Squares show the upper limits of $1.3$ mm fluxes.
The dotted line is the detection limit, $2.5$ mJy, of the observation by
Duvert et al. (2000). 
The distance $D=140$ pc is assumed.
\label{fig:emission}
}
\end{figure}

Evolution of the dust density causes attenuation of thermal
emission from the disk.
The thermal flux $F_{\nu}$ observed at a frequency $\nu$ is (Adams,
Lada, \& Shu 1988)
\begin{equation}
4 \pi D^2 F_{\nu} = 4  \pi \int_{r_{\rm in}}^{r_{\rm out}} B_{\nu}
(T) [1 - \exp (-\tau_{\nu})] 2 \pi r dr \ ,
\end{equation}
where $D$ is the distance of the disk from the observer, $B_{\nu}$ is the
Planck function, $\tau_{\nu}$ is the vertical optical depth of the disk.
For simplicity, we assumed that the disk is observed face-on from a
distance of $D=140$ pc.
Because most of the millimeter flux comes from the optically thin, outer
part of the disk, the viewing angle does not affect the flux
significantly, unless the disk is close to edge-on.
The optical depth is $\tau_{\nu}=\kappa_{\nu} \Sigma_d /f_{\rm dust}$,
where the opacity $\kappa_{\nu}$ is defined for unit mass of the initial
disk gas that is well mixed to the dust (with the dust-to-gas ratio
$f_{\rm dust}=0.01$), and thus $\kappa_{\nu} / f_{\rm dust}$ is the
opacity for unit dust mass.
We assume that the dust opacity index is $\beta=1$ and $\kappa_{\nu} =
0.1 (\nu / 10^{12} \ {\rm Hz} )$, which is observationally suggested
(Beckwith \& Sargent 1991; Kitamura et al. 2002).
The millimeter flux is nearly proportional to the disk mass.
We assume an initial disk mass $M_g = 2.5 \times 10^{-2} M_{\sun}$.
Figure \ref{fig:emission} shows the time evolution of the disk emission
at 1.3 mm wavelength, compared to observations by Beckwith et
al. (1990) and Osterloh \& Beckwith (1995).
In $A=1$ model, the disk emission quickly decays and in $6 \times 10^5$
yr it becomes lower than the detection limit.
To keep the thermal emission over $10^6$ yr, the dust grains must be fluffy
or the gas disk must be massive, as shown by the result for $A=0.1$ model.
(In cases of massive disks, the magnitude of the millimeter flux would
be larger than the line of $A=0.1$ in Fig. \ref{fig:emission}, which is
drawn for $M_g = 2.5 \times 10^{-2} M_{\sun}$.)
Even in the fluffy dust or massive disk cases, however, grain growth
from 1 mm to 1 cm recovers $A=1$, and then the emission decays in
$6 \times 10^5$ yr.

\subsection{High Viscosity Models \label{sec:highvis}}

\begin{figure}
\epsscale{1.0}
\plotone{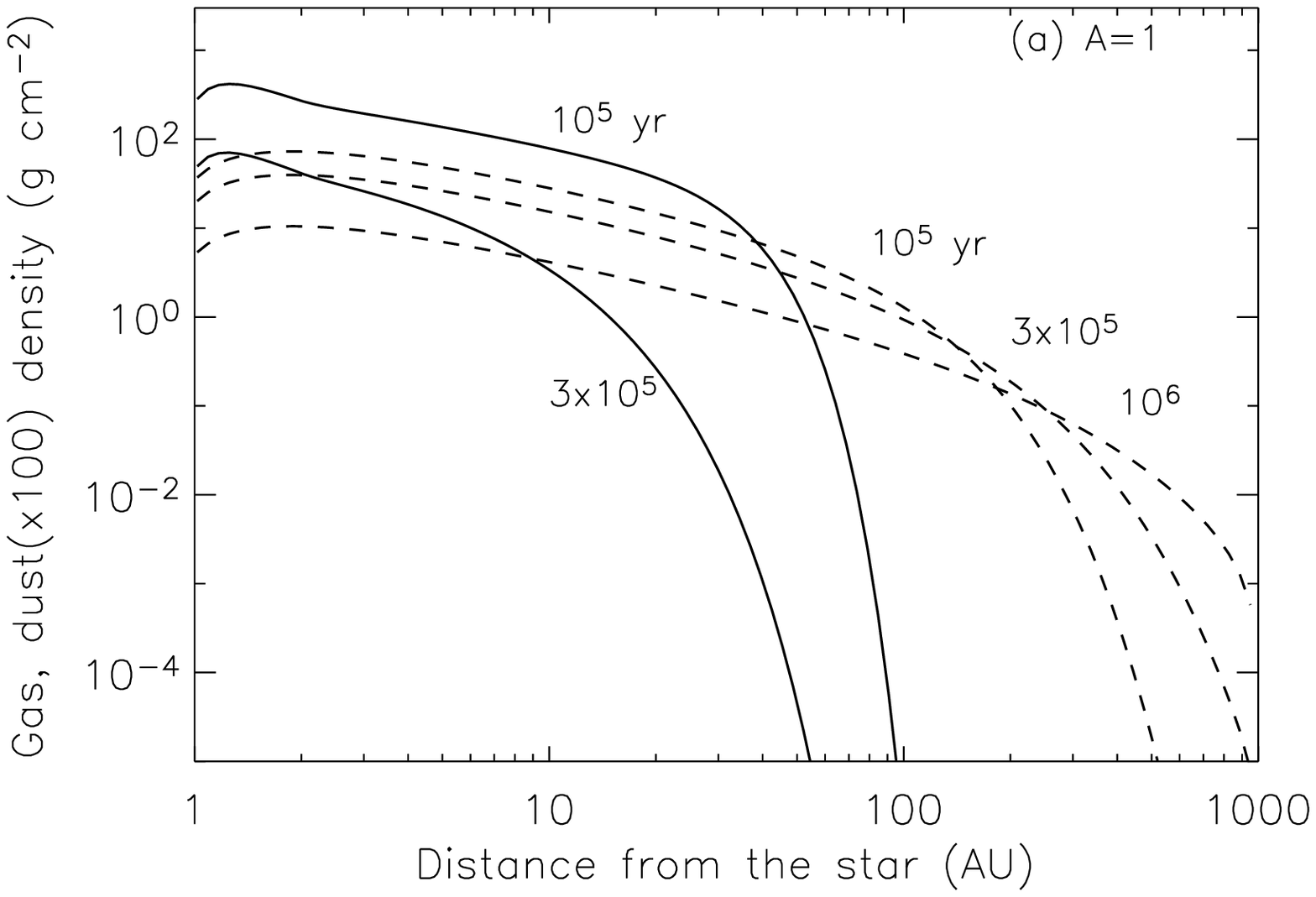}
\plotone{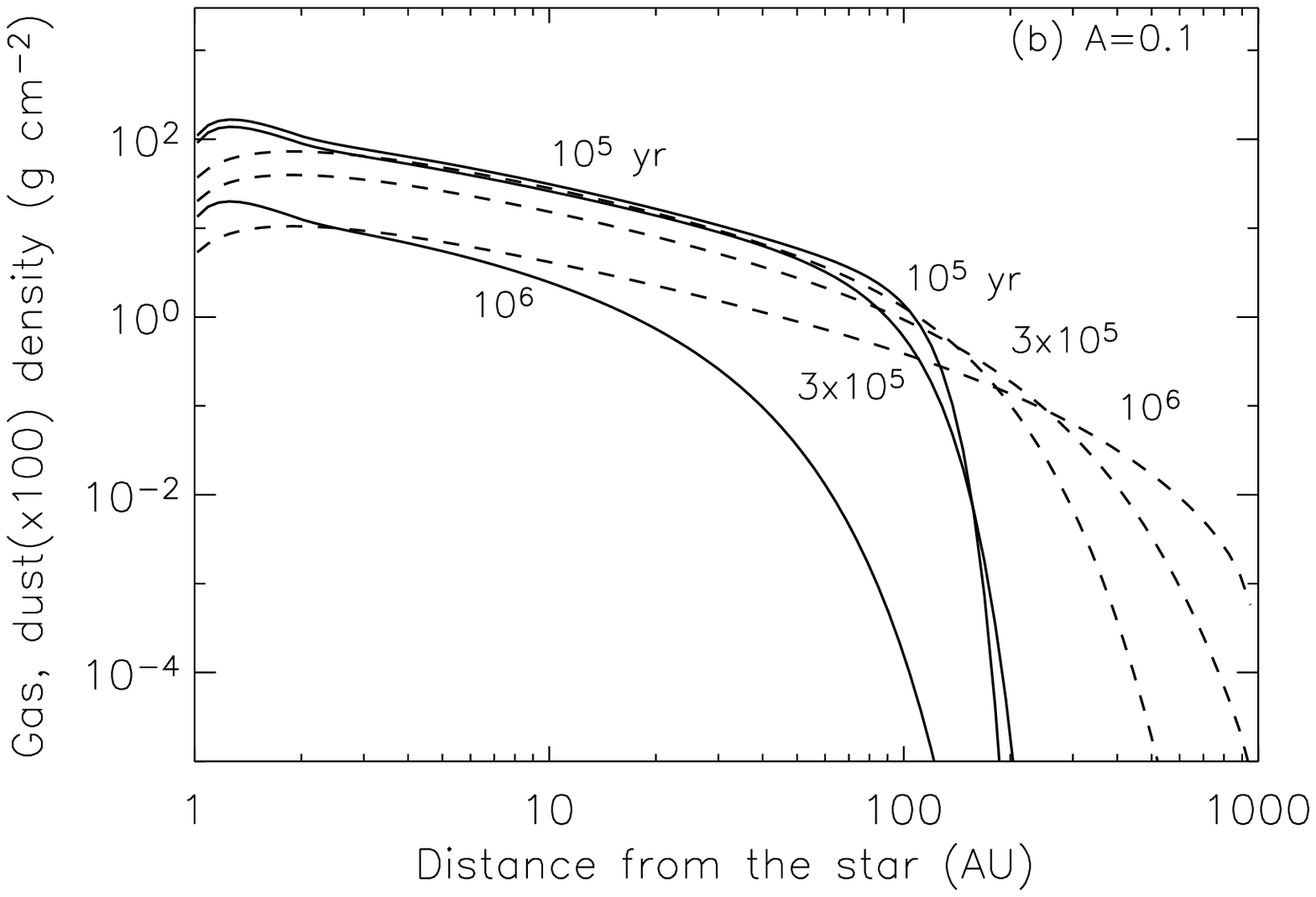}
\caption{
Same as Fig. \ref{fig:denevo}, but for a high viscosity disk
($\alpha=10^{-2}$). 
(a) $A=1$.
At $10^6$ yr, the dust surface density is below the $y$-range of the
figure.
(b) $A=0.1$.
\label{fig:denevo_hv}
}
\end{figure}

Figure \ref{fig:denevo_hv} shows the dust density evolution of a disk
with a viscosity parameter $\alpha = 10^{-2}$.
The gas disk suffers rapid viscous accretion and in $10^6$ yr the gas
density within 100 AU reduces significantly.

\begin{figure}
\epsscale{1.0}
\plotone{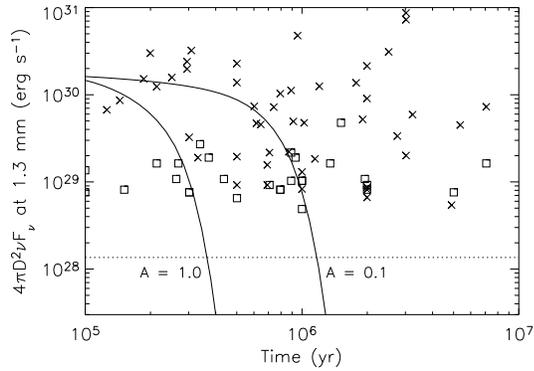}
\caption{
Same as Fig. \ref{fig:emission}, but for a high viscosity disk
($\alpha=10^{-2}$).
\label{fig:emission_hv}
}
\end{figure}

On the other hand, because the dust drift velocity (the second term in
eq. [\ref{eq:dustvel2}]) is independent of $\alpha$, 
evolution of the dust disk does not differ significantly
from the models with low viscosity ($\alpha = 10^{-3}$), except that the
dust accretion is slightly faster.
In $A=0.1$ model shown in Figure \ref{fig:denevo_hv}$b$, the dust migration
time-scale at the initial state ($t=0$) is comparable to the gas
lifetime of $10^6$ yr.
As the gas density declines, however, the dust
accretion is sped up (see eq. [\ref{eq:dustvel2}]).
We have thus a faster dust accretion than in the low viscosity case.
The dust always disappears before the gas dispersal.
Evolution of the millimeter emission shown in Figure
\ref{fig:emission_hv} is similar to the low viscosity models, but is
a few times faster.

\subsection{Meter Sized Bodies \label{sec:largepar}}

\begin{figure}
\epsscale{1.0}
\plotone{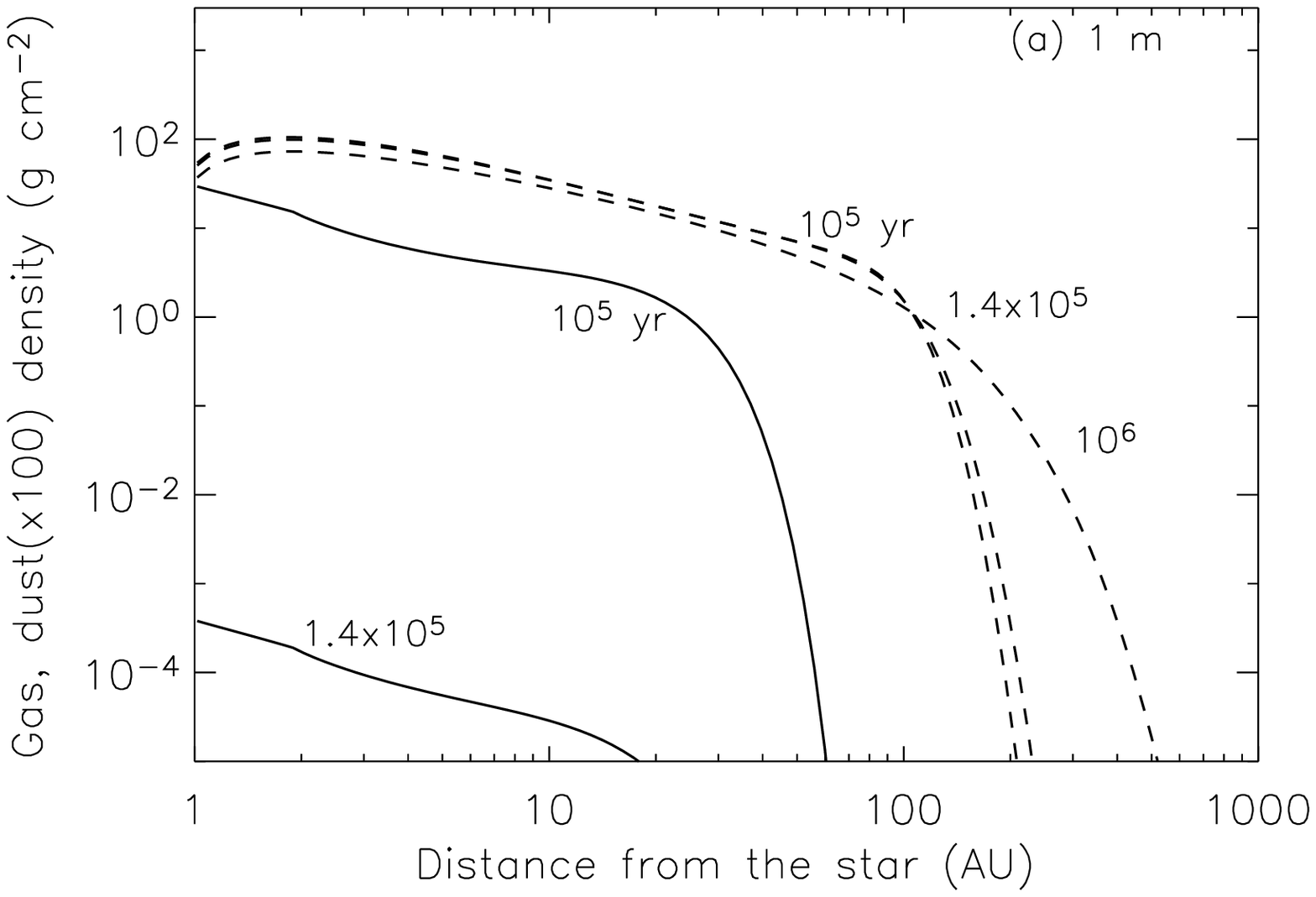}
\plotone{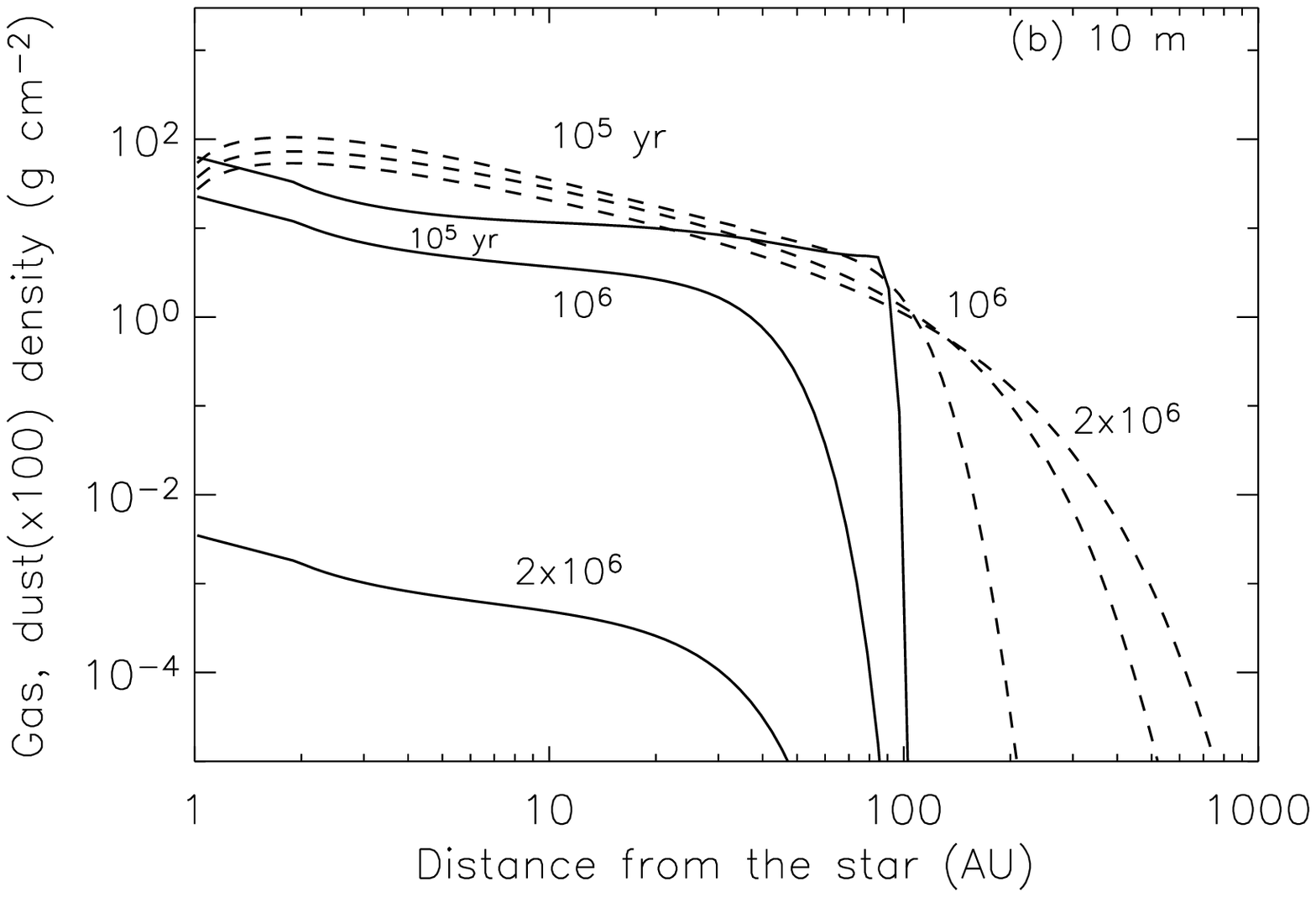}
\caption{
Same as Fig. \ref{fig:denevo}, but for large bodies.
(a) $s=1$ m, $\rho_p=1 \ {\rm g \ cm}^{-3}$.
(b) $s=10$ m.
\label{fig:denevo_lgp}
}
\end{figure}

We consider a case in which the largest bodies have grown to meter
size.
The non-dimensional stopping time of meter sized bodies is larger
than unity and the
drift velocity is smaller for larger bodies as shown in Figure
\ref{fig:driftvel}.
Thus, there is a lower limit for the body size to survive in a 
disk for $10^6$ yr.
Figure \ref{fig:denevo_lgp} shows the evolution of the dust disks for 1 m
and 10 m bodies.
The gas mean free path is less than 1 m in the inner disk of $r \la 4$ AU,
and less than 10 m for $r \la 10$ AU.
Epstein's drag law is thus not valid in the innermost part of the
disk, but we neglect the divergence from the Epstein's law because our
interest is mainly on the outer part.
For bodies of $T_s \gg 1$, the gas drag does not cause any
significant diffusion of the dust, so we set $F_{\rm dif}=0$ in
equation (\ref{eq:cont_dust}).
Dust concentration at the midplane does not change the gas velocity 
profile significantly.
The non-dimensional stopping time of the gas is $T_{s,g} = (\rho_g / \rho_d)
T_s$. 
In order for the gas velocity profile to be affected by the dust, $T_{s,g}$
must be less than unity, i.e., $\rho_d / \rho_g > T_s$. 
At 50 AU, bodies of 1 m have $T_s \simeq 20$, thus $T_{s,g} < 1$ requires
that the dust layer's thickness is less than $f_{\rm dust} T_s^{-1} \sim
5 \times 10^{-4}$ of the gas disk thickness.
We assume that the turbulence of the gas disk is strong enough to prevent such
a thin dust layer.
The particle physical density is $\rho_p = 1 \ {\rm g \ cm}^{-3}$, assuming
that bodies as large as 1 m are compacted.
 
Bodies of 1 m cannot survive for $10^6$ years, as shown in Figure 
\ref{fig:denevo_lgp}$a$.
Dust grains have to grow up to at least $\sim 10$ m in order to reside in
the disk more than $10^6$ yr.
However, bodies larger than $10$ m cannot emit millimeter continuum
effectively.
Figure 4 of Miyake \& Nakagawa (1993) shows that if most of the dust
mass is confined in 10 m bodies, the emissivity at a wavelength of 1 mm is
three orders of magnitude less than the value for 1 mm grains.

In conclusion, the lifetime of dust disks is controlled by the dust
migration time-scale, although a gas disk with a low viscosity
($\alpha \la 10^{-3}$) can survive much longer than the dust disk.
A dust disk composed of 1 mm compact ($\rho_p \ga 1 \ {\rm g \
  cm}^{-3}$) grains in a low mass ($M_g \la 2.5\times 10^{-2}
M_{\sun}$) gas disk cannot survive for $10^6$ yr.
We need fluffy ($\rho_p \la 0.1 \ {\rm g \ cm}^{-3}$) grains or a
massive ($M_g \ga 2.5\times 10^{-1} M_{\sun}$) gas disk in order to
explain the observed dust emission around CTTSs of ages $\ga 1$ Myr.
Survival of 1 cm grains is much harder than 1 mm grains.


\section{Constraint on the Dust Growth Time \label{sec:growth}}

As shown in \S\ref{sec:lowvis}, dust grains of 1 cm hardly survive
the rapid accretion for $10^6$ yr, while observations suggest that 
grains around some CTTSs as old as $10^6-10^7$ yr are larger than 1
mm.
This implies one of the following possibilities:
(1) A large enough amount of 1 mm grains needed to explain the observed
millimeter flux is conserved for more than $10^6$ yr without growing to
1 cm.
Some part of the dust may have grown up to 1 cm and been removed
from the disk.
As the number density of 1 mm grains decreases, the growth time-scale
becomes longer and the residual 1 mm grains keep their small sizes for
more than $10^6$ yr.
(2) When grains grow beyond a critical size, at which $T_s=1$, the
radial velocity decreases as their sizes grow, as seen from equation
(\ref{eq:dustvel1}). 
The dust grains around CTTSs have already grown large enough ($\sim
10$ m) to be decoupled from the gas and keep their orbits steady for
$10^6$ yr.
The grains must grow quickly enough to avoid a large migration at the
critical size.
In addition, the optical depth of the disk must be large enough to
explain the observed millimeter flux.
Because only bodies larger than $\sim 10$ m can reside in the disk
for $10^6$ yr and such large bodies cannot emit millimeter continuum
effectively, millimeter sized grains must be produced from the larger
bodies.
(3) The rapid accretion of grains removes most of the dust from the
disk in a time-scale less than $10^6$ yr, but new dust is continuously
replenished from the outside, such as an envelope that surrounds the
disk.

If the possibilities (1) or (2) are the case, we can derive a constraint
on the growth time-scale of the dust.
In this section, we consider the dust growth time-scale using a simple 
analytical estimation.

\subsection{Growth Time-scale}

Dust grains collide with other grains and stick together to grow.
We assume that growth of the largest grains occurs mainly through
collisions among the largest grains, because the dust mass is
dominated by them, as discussed in \S\ref{sec:th_growth}.
We also assume that the grains well couple to the gas ($T_s <1$).
The growth of the largest grain is written as
\begin{equation}
\frac{d m_d}{dt} = \frac{d}{dt} \left( \frac{4}{3} \pi s^3 \rho_p
\right) = C_{\rm stk} \pi (2 s)^2 \rho_d \Delta v \ ,
\label{eq:groweq}
\end{equation}
where the sticking parameter $C_{\rm stk}$ is the probability of
coagulation per collision, $\rho_d$ is the dust mass density
at the midplane, and
$\Delta v$ is the relative velocity of collisions.
The collisional cross section of equal-sized grains is $\pi (2 s)^2$.
The sticking parameter is unknown and we discuss the constraint on it.

The collision velocity comes from the turbulent motion, the radial
drift, and the sedimentation of grains.
The relative velocity of two grains in the turbulent gas is $\Delta
v_t \simeq v_{\rm turb} \sqrt{T_s} \simeq \sqrt{\alpha T_s} c$ for
grains of $T_s < 1$  (V\"olk et al. 1980), where we assume that the
largest eddies have a typical velocity $v_{\rm turb}
\simeq \sqrt{\alpha} c$ and a typical turnover time $\Omega_{\rm K}$
(Cuzzi et al. 2001). 
The relative velocity due to radial drift is $\Delta v_d \simeq \eta
T_s v_{\rm K}$ (see eq. [\ref{eq:dustvel1}]), and that due to
  sedimentation is $\Delta v_z \simeq (z/r) T_s v_{\rm K}$ (eq. [14] of
  Takeuchi \& Lin 2003).
For grains with small $T_s$, the turbulent motion controls the collision
velocity, and $\Delta v \simeq \Delta v_t$.
In the equilibrium state of the vertical density distribution of the dust, the
grain sedimentation is balanced by the turbulent diffusion.
Hence, the relative velocity due to the sedimentation, $\Delta v_z$, is
similar to or smaller than that from the turbulence and can be
neglected ($\Delta v_z \simeq (h_d/r) T_s v_{\rm K} \la \sqrt{2
  \alpha T_s} c \simeq \Delta v_t$, where $h_d$ is the scale
height of the dust disk and eq. [\ref{eq:hd}] below is used.)
The condition for $ \Delta v_t > \Delta v_d$ is
\begin{equation}
T_s < \frac{\alpha h_g^2}{\eta^2 r^2} \sim \alpha \left( \frac{h_g}{r}
\right)^{-2} \sim 0.1 \ ,
\end{equation}
where equation (\ref{eq:eta2}), $\alpha=10^{-3}$ and $(h_g/r) \sim 0.1$
is used.
For grains of $T_s \simeq 1$, the collision velocity is determined by
the radial drift, and $\Delta v \simeq \Delta v_d$.

The midplane dust density is estimated by 
\begin{equation}
\rho_d = \frac{\Sigma_d}{\sqrt{2 \pi} h_d} \ ,
\label{eq:dustden}
\end{equation}
where the same relation between the surface and midplane gas densities of 
equation (\ref{eq:midgasden}) is applied to the dust.
If dust grains are small enough to be well mixed to the gas, i.e., in
the limit of small $T_s$, then $h_d=h_g$.
For relatively large grains, the dust sediments to the
midplane and the thickness of the dust disk is smaller than that of the
gas.
As derived in equation (34) in Paper I (with the Schmidt number ${\rm
Sc}=1$), the dust disk scale height becomes 
\begin{equation}
h_d = \left[ 2 \ln \left( \frac{\alpha}{T_s} +1 \right) \right]^{1/2} h_g
\approx  \left( \frac{2 \alpha}{T_s} \right)^{1/2} h_g \ ,
\label{eq:hd1}
\end{equation}
for $T_s \ga \alpha$.
Because $h_d$ cannot be larger than $h_g$, the approximated scale height
is expressed as
\begin{equation}
h_d = \min \left[1,  \left( \frac{2 \alpha}{T_s} \right)^{1/2} \right]
h_g \ ,
\label{eq:hd}
\end{equation}
which is applicable for $T_s < 1$.

The growth time-scale is defined as an $e$-folding time in size.
From equations (\ref{eq:groweq}), (\ref{eq:dustden}), and (\ref{eq:hd}),
it becomes
\begin{equation}
\tau_{\rm grow} = \frac{s}{ds/dt} = \frac{\sqrt{2 \pi} s \rho_p h_g}{C_{\rm
  stk} \Sigma_d \Delta v} \min \left[ 1, \ \left( \frac{2 \alpha}{T_s}
  \right)^{1/2} \right] \ .
\label{eq:gtime1}
\end{equation}
For grains that have $2 \alpha < T_s < \alpha (r/h_g)^2$, the growth
time-scale reduces to 
\begin{equation}
\tau_{\rm grow} = \frac{4 \Sigma_g}{\sqrt{\pi} C_{\rm stk} \Sigma_d}
  \Omega_{\rm K}^{-1} \ ,
\label{eq:gtime2}
\end{equation}
where $\Delta v \simeq \Delta v_t \simeq \sqrt{\alpha T_s} c$ and
equations (\ref{eq:gasheight}) and (\ref{eq:stoptime2}) are used.
In this range of $T_s$, the growth time is
independent of the grain size $s$.

\subsection{Slow Growth}

We first consider the possibility that grains keep their sizes less
than 1 cm to avoid rapid accretion for more than $10^6$ yr.

\begin{figure}
\epsscale{1.0}
\plotone{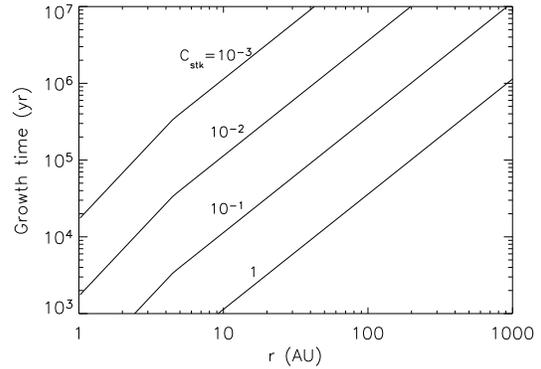}
\caption{
Dust growth time-scale, $\tau_{\rm grow}$, for various sticking
parameters $C_{\rm stk}$.
The disk is at the initial state ($t=0$) of the fiducial model described in
\S\ref{sec:diskmodel}.
\label{fig:growth}
}
\end{figure}

Grains of 1 cm with $\rho_p= 0.1 \ {\rm g \ cm}^{-3}$ satisfy the
condition 
$2 \alpha < T_s < \alpha (r/h_g)^2$ if they are in $5 \ {\rm AU} \la r \la
200$ AU of a disk with $\alpha=10^{-3}$ and $(h_g/r) \sim 0.1$ (see
the $A=1$ line in Fig. \ref{fig:stoptime}).
For such grains, the growth time-scale is calculated by equation
(\ref{eq:gtime2}).
Inside 5 AU, it is calculated by equation (\ref{eq:gtime1}) with $\Delta
v \simeq \Delta v_d$. 
Figure \ref{fig:growth} shows the growth time-scale for various values 
of the sticking parameter $C_{\rm stk}$.
If grains stick effectively ($C_{\rm stk} \ga 0.1$), the
growth time is less than $10^6$ yr in most part of the disk ($r \la
200$ AU).
It is thus expected that grains rapidly grow up to 1 cm and are
quickly removed.
If the sticking parameter is $C_{\rm stk} = 10^{-2}$, dust growth
in the outer disk ($r \ga 40$ AU) takes more than $10^6$ yr and the outer
dust disk can survive.

In order to see the effect of dust growth, we put a sink term,
\begin{equation}
\dot{\Sigma}_d = - \frac{\Sigma_d}{\tau_{\rm grow}} \ ,
\end{equation}
in the right hand side of equation (\ref{eq:cont_dust}).
In this treatment, we assume that on a time-scale $\tau_{\rm grow}$ the dust
grains grow from 1 mm to 1 cm and are removed from the disk.
The growth time-scale $\tau_{\rm grow}$ is the $e$-folding time in the size, and
the actual growth time from 1 mm to 1 cm is probably longer by a factor,
$\ln 10 \approx 2.3$, but we neglect the difference of this factor.

\begin{figure}
\epsscale{1.0}
\plotone{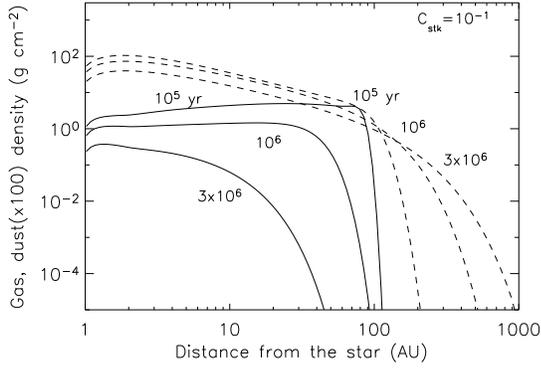}
\caption{
Same as Fig. \ref{fig:denevo}, but with dust removal through its growth.
The sticking probability is $C_{\rm stk}=0.1$. 
The dust parameters are $s=1$ mm, $\rho_p=0.1 \ {\rm g \ cm}^{-3}$.
\label{fig:denevo_gr}
}
\end{figure}

\begin{figure}
\epsscale{1.0}
\plotone{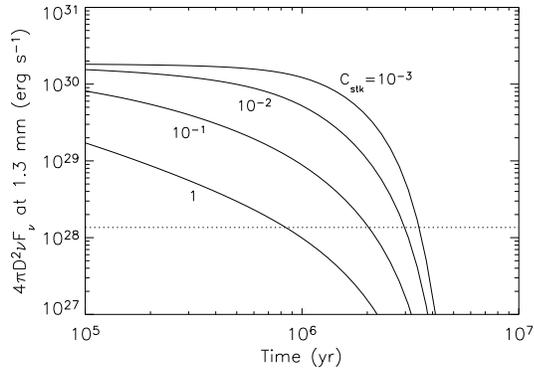}
\caption{
Same as Fig. \ref{fig:emission}, but with dust removal through its growth.
The sticking parameters are $C_{\rm stk}=10^{-3}$, $10^{-2}$, $10^{-1}$,
and 1, respectively, from the upper line.
\label{fig:emission_gr}
}
\end{figure}

Figure \ref{fig:denevo_gr} shows the density evolution of the dust with
the sticking parameter $C_{\rm stk} = 10^{-1}$. 
The grains first grow quickly with a much shorter time-scale than the
migration time-scale.
As the dust density decreases, the growth time-scale increases
(see eq. [\ref{eq:gtime2}]) and finally becomes as large as the age of
the disk. (The dust growth itself cannot enlarge the growth time beyond
the age of the disk.)
In $10^5$ yr, the growth time-scale has become similar to the age
($10^5$ yr) in the entire dust disk.
At this stage, the growth time is almost constant with $r$.
From equation (\ref{eq:gtime2}), a constant growth time means that
the surface density profile of the dust approaches $\Sigma_d \propto
\Sigma_g / \Omega_{\rm K} \propto r^{1/2}$.
At $t \sim 10^6$ yr, the growth time increases to $10^6$ yr, which is
comparable to the dust migration time-scale.
After that, grain migration affects the shape of the density profile.
Figure \ref{fig:emission_gr} shows the time evolution of the dust thermal
emission for various values of the sticking parameter $C_{\rm stk}$.
If dust growth is effective and $C_{\rm stk} \approx 1$, the dust disk
cannot survive for $10^6$ yr and the thermal emission quickly decays.
To explain the observed dust emission, the sticking parameter must be
less than $\sim 0.1$.

The slow growth scenario may cause a difficulty in forming planets at
large distances from the star ($\ga 40$ AU).
We assumed that grains in the outer part of the disk keep their sizes
smaller than 1 cm for $10^6$ yr.
After $10^6$ yr, the grains have grown to 1 cm, however, and they
start rapid accretion and disappear or at least
migrate to the inner part of the disk before growing to planetesimals.
In this scenario, it is difficult to form planets in the outer region of
the disk, though planet formation may be possible in the inner disk.

\subsection{Fast Growth \label{sec:fastgrowth}}

The other possibility is that the dust grains have
already grown large enough to stop their migration.
After the grain size becomes larger than the critical size, at which
$T_s$ is unity, the grains' radial motion slows down as they
grow larger (see eq.[\ref{eq:dustvel1}] and Fig. \ref{fig:driftvel}).
Finally, the grains completely decouple from the gas and stop in the
disk.
Before they grow up to such large bodies, however, they must survive the
period of rapid migration.

\begin{figure}
\epsscale{1.0}
\plotone{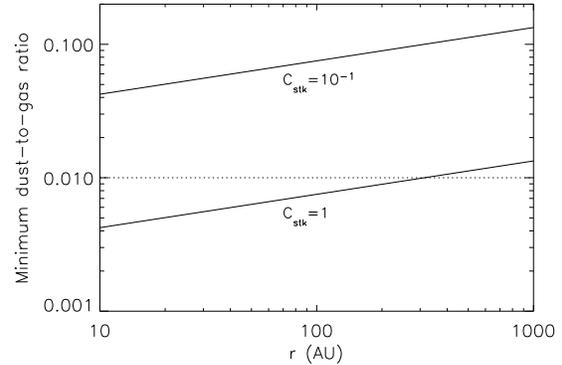}
\caption{
Condition on the dust-to-gas ratio for dust survival by fast growth,
given by equation (\ref{eq:grcond2}).
The disk is in the initial state ($t=0$) of the fiducial model described in
\S\ref{sec:diskmodel}.
The condition, $\tau_{\rm grow} < \tau_{\rm dust,min}$, is satisfied above
the upper solid line for $C_{\rm stk}=0.1$ and above the lower solid
line for $C_{\rm stk}=1$.
The dotted line shows the dust-to-gas ratio $f_{\rm dust} = 10^{-2}$.
The figure shows only $r > 10$ AU, where Epstein's gas drag law
is valid.
\label{fig:grcond2}
}
\end{figure}

When $T_s=1$, the radial velocity reaches the maximum value,
$v_{d, \rm max} = - (\eta / 2) v_{\rm K}$,
and the migration time $\tau_{\rm dust}$ becomes minimum,
\begin{equation}
\tau_{\rm dust, min} = \frac{r}{|v_{d, \rm max}|} = 2 \eta^{-1} \Omega_{\rm
  K}^{-1} \ .
\end{equation}
(The dust disk thickness of $T_s=1$ grains is $\sqrt{2 \alpha} h_g$
as seen in eq. [\ref{eq:hd1}].
The midplane dust density is increased by a factor $(2 \alpha)^{-1/2}$,
but as long as $\alpha > 10^{-4}$, the dust density does not exceed the gas
density and the gas velocity profile is not modified.)
From expression (\ref{eq:stoptime2}) of the non-dimensional stopping time
in Epstein's gas drag law, the grain size at $T_s=1$ is
\begin{equation}
s = \frac{2 \Sigma_g}{\pi \rho_p} \ .
\label{eq:sizecrt}
\end{equation}
(Epstein's law is valid for grains smaller than 10 m at 10 AU and
smaller than 10 km at 100 AU.)
In order to survive rapid migration, the dust must grow quickly and
pass through this hazardous period before falling into the star.
The growth time-scale is, from equations (\ref{eq:gtime1}) and
(\ref{eq:sizecrt}) with $ \Delta v \simeq v_{d, \rm max}$ and $T_s=1$,
\begin{equation}
\tau_{\rm grow} = \frac{4 \sqrt{\alpha} \Sigma_g h_g}{\sqrt{\pi} C_{\rm
  stk} \Sigma_d v_{d, \rm max}} \ .
\end{equation}
The condition $\tau_{\rm grow} < \tau_{\rm dust, min}$ becomes
\begin{equation}
\frac{\Sigma_d}{\Sigma_g} > \frac{4 \sqrt{\alpha}}{\sqrt{\pi} C_{\rm
    stk}} \frac{h_g}{r}\ .
\label{eq:grcond2}
\end{equation}
Hence, the dust-to-gas ratio must to be larger than certain values, and
this condition is independent of the dust properties.
Figure \ref{fig:grcond2} shows the values of the right hand side of
equation (\ref{eq:grcond2}) for $C_{\rm stk}=0.1$ and 1
(and $\alpha=10^{-3}$).
The condition  $\tau_{\rm grow} < \tau_{\rm dust, min}$ is satisfied if
the value of $\Sigma_d / \Sigma_g$ is above the solid lines.
If we assume the dust-to-gas ratio is $f_{\rm dust}=\Sigma_d /
\Sigma_g=10^{-2}$, the sticking parameter $C_{\rm stk}$ must be close to
unity in order for the dust to grow fast enough (at $r \la 300$ AU).
If the grains are as less sticky as $C_{\rm stk}=0.1$, the migration
dominates the growth. 

After the grains pass the fast migration period, they still continue to
grow.
In order that grains may reside in the disk for $10^6$ yr, their sizes 
must be larger than $\sim 10$ m, as shown in \S\ref{sec:largepar}.
However, observed millimeter continuum is probably not emitted from such
large bodies.
Thus, grains of $\sim 1$ mm need to be continuously replenished
through collisions of the large bodies, or from an envelope covering the
disk.

One possibility of accounting for the grain growth and the millimeter
flux at the same time is assuming a high value of the dust-to-gas ratio.
It is possible to keep the millimeter flux as high as the
observed values of some CTTSs for $10^6$ yr, if the sticking
probability is $C_{\rm stk}=0.1$ (see Fig. \ref{fig:emission_gr}).
It is also possible for grains inside 20 AU to survive the rapid
accretion phase, if the dust-to-gas ratio is as large as 0.05, for the
same sticking probability $C_{\rm stk}=0.1$ (see
Fig. \ref{fig:grcond2}).
Hence, if the dust-to-gas ratio is several times higher than the solar
abundance and if the sticking probability is of order of 0.1,
then grains of millimeter size in the outer part of the disk can
continue to emit sufficient millimeter continuum, and grains in the inner
disk ($\la 20$ AU) can grow fast enough to survive the rapid accretion
phase.
Calculations including both grain growth and radial
migration are needed to examine this possibility.


\section{Summary}

Motivated by radio observations suggesting grain growth to
1 mm or larger in CTTS disks, we studied the evolution of dust disks
whose main components are millimeter sized grains.

1. Gas drag on millimeter sized grains accelerates the dust disk
accretion to the star.
When grains grow to $\sim 1$ cm, they cannot reside in the disk for over
$10^6$ yr, and as the dust disk accretes, its millimeter continuum emission
decreases.
The lifetimes of grains are longer for a smaller particle physical
density (i.e., fluffy), and in a more massive gas disk.

2. If the observed millimeter continuum of CTTSs comes from millimeter
sized grains, a sufficient amount of grains in this size range must exist more
than $10^6$yr in their disks.
This suggests one of the following possibilities.
One is that millimeter sized grains do not grow rapidly, and it takes more
than $10^6$ yr to become larger than 1 cm.
If the grains' sticking probability is less than $\sim 0.1$, this condition
is satisfied in the outermost region of the disk ($r \ga 100$ AU).
Another possibility is that the grains have already grown to $10$ m or 
larger and they do not migrate rapidly anymore, and millimeter sized 
grains are continuously replenished through collisions of the large bodies.
If the disk's dust-to-gas ratio is $\sim 0.1$ and the sticking probability
is $C_{\rm stk} \sim 0.1$, the observable millimeter continuum holds for
more than $10^6$ yr, emitted mainly from grains of $\sim 1$ mm in
the outer disk, and in the inner disk the grains can grow larger than 10
m before falling to the star.
The other possibilities are that millimeter sized grains are replenished
from a surrounding envelope, or that the small $\beta$ of the observed 
millimeter continuum is not due to dust growth.
In order to examine these possibilities, we need to perform simulations 
taking account of both dust migration and dust growth at the same time.

\acknowledgements
We gratefully acknowledge useful discussions with Cathie Clarke, which
drew our attention to this problem.
We are also grateful to Sozo Yokogawa and Yoichi Itoh for useful
comments, and to Anthony Toigo for careful reading of the manuscript.
We acknowledge helpful comments from the referee.
This work was supported in part by an NSF grant AST 99 87417 and in part
by a special NASA astrophysical theory program that supports a joint
Center for Star Formation Studies at UC Berkeley, NASA-Ames Research
Center, and UC Santa Cruz.
This work was also supported by NASA NAG5-10612 through its Origin
program, JPL 1228184 through its SIM program, and the 21st Century COE
Program of MEXT of Japan, ``the Origin and Evolution of Planetary
Systems.''



\clearpage


\begin{thebibliography}{}

\bibitem[Adachi et al. 1976]{ada76} Adachi, I., Hayashi, C., \&
Nakazawa, K. 1976, Prog. Theor. Phys. 56, 1756

\bibitem[Adams, Lada \& Shu 1988]{ada88} Adams, F. C., Lada, C. J., \& Shu,
F.H. 1988, \apj, 326, 865

\bibitem[Armitage, Clarke, \& Palla 2003]{arm03} Armitage, P. J., Clarke, 
C. J., \& Palla, F. 2003, \mnras, 342, 1139

\bibitem[Bath \& Pringle,1981]{bat81} Bath, G. T., \& Pringle, J. E. 1981,
\mnras, 194, 967

\bibitem[Beckwith, Henning, \& Nakagawa 2000]{bec00} Beckwith, S. V. W., 
Henning, T., \& Nakagawa, Y. 2000, in Protostars and Planets IV, eds. 
Mannings, V., Boss, A.P., \& Russell, S. S. (Tucson: University of Arizona
Press), 533 

\bibitem[Beckwith \& Sargent 1991]{bec91} Beckwith, S. V. W., \&
Sargent, A. I. 1991, \apj, 381, 250

\bibitem[Beckwith et al. \ 1990]{bec90} Beckwith, S. V. W., Sargent, A.
I., Chini, R. S., \& G\"usten, R. 1990, \aj, 99, 924

\bibitem[Cuzzi et al. 2001]{cuz01} Cuzzi, J. N., Hogan, R. C., Paque, J. M.,
 \& Dobrovolskis, A. R. 2001, \apj, 546, 496

\bibitem[Dutrey et al. 1996]{dut96} Dutrey, A., Guilloteau, S., Duvert, G.,
Prato, L., Simon, M., Schuster, K., \& Menard, F. 1996, \aap, 309, 493

\bibitem[Duvert et al. 2000]{duv} Duvert, G., Guilloteau, S., Menard, F., 
Simon, M., \& Dutrey, A. 2000, \aap, 355, 165

\bibitem[Hartmann et al. 1998]{har93} Hartmann, L., Calvet, N.,
Gullbring, E., \& D'Alessio, P. 1998, \apj, 495, 385

\bibitem[Hayashi, Nakazawa, \& Nakagawa 1985]{hay85} Hayashi, C.,
Nakazawa, K., \& Nakagawa, Y. 1985, in Protostars and Planets II, ed. D.
C. Black \& M. S. Matthews (Tucson:Univ. of Arizona Press), 1100

\bibitem[Hellyer 1970]{hel70} Hellyer, B. 1970, \mnras, 148, 383 

\bibitem[Hogerheijde et al. 2003]{hog03} Hogerheijde, M. R., Johnstone,
D., Matsuyama, I., Jayawardhana, R., \& Muzerolle, J. 2003, \apjl, 593, L101

\bibitem[Kenyon \& Hartmann 1995]{ken95} Kenyon, S. J., \& Hartmann, L. 1995,
\apjs, 101, 117

\bibitem[Kitamura et al. 2002]{kit02} Kitamura, Y,. Momose, M., Yokogawa, 
S., Kawabe, R., Tamura, M., \& Ida, S. 2002, \apj, 581, 357

\bibitem[Lin \& Papaloizou 1986b]{lin86b} Lin, D. N. C., \& Papaloizou,
J. C. B. 1986, \apj, 309, 846 

\bibitem{lyn74} Lynden-Bell, D., \& Pringle, J. E. 1974, \mnras, 168, 603

\bibitem[Mannings \& Emerson 1994]{man94} Mannings, V., \& Emerson,
J. P. 1994, \mnras, 267, 361

\bibitem[Mathis et al. 1977]{mat77} Mathis, J. S., Rumpl, W., \& Nordsieck,
K.H. 1977, \apj, 217, 425

\bibitem[Miyake \& Nakagawa 1993]{miy93} Miyake, K., \& Nakagawa, Y. 1993,
Icarus, 106, 20

\bibitem[Mizuno et al. 1988]{miz88} Mizuno, H., Markiewicz, W. J., \& V\"olk,
H. J. 1988, \aap, 195, 183

\bibitem[Nakagawa, Sekiya \& Hayashi]{nak86} Nakagawa, Y.,
Sekiya, M., \& Hayashi, C. 1986, Icarus, 67, 375

\bibitem[Ohashi et al. 1991]{oha91} Ohashi, N., Kawabe, R., Ishiguro,
M., \& Hayashi, M. 1991, \aj, 102, 2054

\bibitem[Ohashi et al. 1996]{oha96} Ohashi, N., Hayashi, M., Kawabe, R.,
\& Ishiguro, M. 1996, \apj, 466, 317 

\bibitem[Osterloh \& Beckwith 1995]{ost95} Osterloh, M., \& Beckwith, S. V.
W. 1995, \apj, 439, 288

\bibitem[Press et al. 1992]{pre92} Press, W. H., Flannery, B. P.,
Teukolsky, S. A., \& Vetterling, W. T. 1992, Numerical Recipes in
Fortran (Cambridge University Press)

\bibitem[Schneider et al. 1999]{sch99} Schneider, G., et al. 1999,
\apjl, 513, L127

\bibitem[Shakura \& Sunyaev 1973]{sha73} Shakura, N. I., \& Sunyaev,
R. A. 1973, \aap, 24,337

\bibitem[Takeuchi \& Lin 2002]{tak02} Takeuchi, T., \&
Lin, D. N. C. 2002, \apj, 581, 1344, (Paper I)

\bibitem[Takeuchi \& Lin 2003]{tak03} Takeuchi, T., \&
Lin, D. N. C. 2003, \apj, 593, 524

\bibitem[van de Hulst 1981]{van81} van de Hulst, H. C. 1981, Light
Scattering by Small Particles (New York:Dover)

\bibitem[van Leer 1977]{van77} van Leer, B. 1977, J. Comput. Phys., 23,
276
\bibitem[V\"olk et al. 1980]{vol80} V\"olk, H. J., Morfill, G. E., Roeser,
S., \& Jones, F. C. 1980, \aap, 85, 316

\bibitem[Weidenschilling 1977]{wei77} Weidenschilling, S. J. 1977,
\mnras, 180, 57

\bibitem[Yoshida et al. 2003]{yos03} Yoshida, F., Nakamura, T.,
Watanabe, J., Kinoshita, D., Yamamoto, N., \& Fuse, T. 2003, \pasj, 55,
701 

\end{thebibliography}
\end{document}